\documentclass[hyper]{JHEP} 

\usepackage{epsfig}

\newcommand\fverb{\setbox\pippobox=\hbox\bgroup\verb}
\newcommand\fverbdo{\egroup\medskip\noindent%
			\fbox{\unhbox\pippobox}\ }
\newcommand\fverbit{\egroup\item[\fbox{\unhbox\pippobox}]}
\newbox\pippobox
\title{Tachyon Kink on non-BPS Dp-brane
in the General Background}

\author{by J. Kluso\v{n}\\
     Department of Theoretical Physics and Astrophysics\\

                   Faculty of Science, Masaryk University\\

Kotl\'{a}\v{r}sk\'{a} 2, 611 37, Brno\\

Czech Republic\\

    E-mail: \email{klu@physics.muni.cz}}

\preprint{\hepth{0508239}}

\abstract{This paper is devoted to the
study of the tachyon kink on 
the worldvolume
of a  non-BPS Dp-brane that is embedded in
general background, including $NS-NS$ two form
$B$ and also general Ramond-Ramond field.
We will explicitly show that the dynamics of
the kink is described by the equations of motion
that arrise from the  DBI and WZ action for
D(p-1)-brane.}

\def\bA{\mathbf{A}}

\def\bAi{(\mathbf{A}^{-1})}
\def\mat{\tilde{\mathbf{a}}}

\def\mati{(\tilde{\mathbf{a}}^{-1})}
\def\mF{\mathcal{F}}
\def\mtF{\tilde{\mathcal{F}}} 
\def\mtC{\tilde{C}}
\begin{document}
\section{Introduction and Summary}\label{first}
The study of the open string tachyon
brought significant progress in the understanding
of the nonperturbative aspect of string
theory 
\footnote{For review of the open string
tachyon condensation, see 
\cite{Sen:1999mg,Ohmori:2001am,
Taylor:2002uv,
Taylor:2003gn,Sen:2004nf}.}. Among
many  results that were
obtained in the past
 there is a very interesting
observation that shows that
 some aspects of
the tachyon condensation 
can be correctly captured by the
effective field theory description,
where the tachyon effective action (\ref{acg}), 
describing
the dynamics of the tachyon field
on a non-BPS Dp-brane of type IIA and
IIB theory was proposed in 
\cite{Sen:1999md,Bergshoeff:2000dq,
Garousi:2000tr,Kluson:2000iy}
\footnote{For recent discussion of the
effective field theory description of
the tachyon condensation, see
\cite{Sen:2002qa,Fotopoulos:2003yt,
Sen:2003bc,Kutasov:2003er,
Niarchos:2004rw}.}. 

One of the  well known
solutions of the tachyon effective field
theory is a kink solution
which is  supposed to describe
a BPS D(p-1)-brane  
\cite{Sen:2003tm,Kim:2005he,Kim:2003in,
Kim:2003ma,Banerjee:2004cw,Bazeia:2004vc,
Copeland:2003df,Brax:2003rs}. Very
nice analysis of the kink solution
was performed in the
 paper \cite{Sen:2003tm} where it
was shown that the energy density 
of the kink in the effective field
theory is localised on codimension
one surface as in the case of
a BPS D(p-1)-brane. It was then also
shown that the worldvolume theory
of the kink solution is also given
by the Dirac-Born-Infeld (DBI) action
on a BPS D(p-1)-brane. Thus result
demonstrates
 that  the tachyon effective action
reproduces the low energy effective
action on the world-volume of the
soliton.

In our recent paper 
\cite{Kluson:2005hd} we
have extended this analysis to the
spatial dependent tachyon condensation
on a unstable Dp-brane moving in
nontrivial  background  with 
the diagonal form of the metric
\footnote{The similar problem 
was previously studied in 
\cite{Kim:2005he,Kluson:2004yk,
Kluson:2005qx}.}.
We have shown that this form 
of the tachyon condensation
leads to an emergence
of a D(p-1)-brane where the scalar
modes that propagate on the kink 
 worldvolume are solutions
of  the equations of
motion that arise from the DBI action
for D(p-1)-brane that is moving
in  given background and that is 
localised at the core of the kink.

The purpose of this paper is to 
extend this analysis to the general
background, including $NS-NS$  two form
field and Ramond-Ramond forms as well. 
We will study this problem in 
two ways. In the first one we will
consider a non-BPS Dp-brane 
action where the  worldvolume diffeomorphism
is not fixed at all. 
 The analysis of the equation
of motion in this way is  straightforward and
in some sense demonstrates the 
efficiency of the  study of
the Dp-brane dynamics without imposing
any gauge fixing conditions
\footnote{We thank prof. U. Lindstr\"{o}m
for stressing this point to us.}. 
More precisely, we will show that
the spatial dependent tachyon condensation
leads to an emergence of a
D(p-1)-brane whose dynamics is
governed by equation of motion
that arise from the DBI and WZ
action for D(p-1)-brane.
 We will also
show that the mode that characterises
the core of the kink does not depend on
the worldvolume coordinates of the kink
and that all its values are equivalent. 
This result is consistent with the fact
that we do not presume any relations
between worldvolume coordinates and
target space ones so that all positions
of the kink on the worldvolume of a 
unstable Dp-brane are equivalent. 

In the second approach we use the
diffeomorphism invariance so that
we will  presume
that the worldvolume coordinate 
that parametrises the spatial dependent
tachyon condensation is equal to 
one spatial  coordinate in target spacetime.
 We will
then demonstrate that 
 the dynamics of the kink
solution is governed by the equation
of motion of D(p-1)-brane even
if the analysis of these equations 
is more difficult. We will also show
that the mode that describes location
of the kink on the worldvolume
of a non-BPS Dp-brane
 has physical meaning as
the embedding coordinate in
the spatial direction that coincides
with the worldvolume direction. We will
also show that this mode obeys
the equation of motion that arises
from the DBI and WZ term for 
D(p-1)-brane moving in given  background.

These results explicitly demonstrate
that the tachyon like DBI action 
together with WZ term allows
 correct description
of the emergence of a BPS
D(p-1)-brane.  We also hope that
this analysis could be extended to
another situations where the effective
field theory description of the tachyon
condensation could be useful.  
For example, we would like to 
apply  this analysis to the
supersymmetric version of a non-BPS
Dp-brane in general background, 
following again \cite{Sen:2003tm}. 
It would be also nice  to find 
solution of the tachyon equation of motion
that describes D-branes with codimensions
larger then one. In other words,
we would like to see whether we can
describe an emergence
of a D(p-2)-brane that, by definition is
unstable and  hence the tachyon should
be present on the worldvolume of
the kink.  

The rest  of this paper is organised 
as follows. In the next section
(\ref{second}) we will analyse 
the equation of motion for non
-BPS Dp-brane in curved background
without any static gauge presumption. 
We will see that the modes living on the
worldvolume of the kink solve the
equation of motion that arise from the
DBI and WZ action for BPS D(p-1)-brane. 
We will also calculate the stress energy tensor
and we will show that it is equal to the
stress energy tensor for D(p-1)-brane. 
In section (\ref{third}) we will
study the same problem where now
we partially fix the gauge.  We will again
show that the dynamics of the kink is governed
by the DBI and WZ action for BPS D(p-1)-brane. 
\section{Non-BPS Dp-brane
in general background}\label{second}
As in our previous paper we begin
with the Dirac-Born-Infeld like tachyon
effective action in general background
\cite{Sen:1999md,Bergshoeff:2000dq,
Garousi:2000tr,Kluson:2000iy}
\begin{eqnarray}\label{acg}
S=-\int d^{p+1}\xi
e^{-\Phi}V(T)\sqrt{-\det \bA} \ ,
\nonumber \\
\bA_{\mu\nu}=g_{MN}
\partial_\mu X^M\partial_\nu X^N+
b_{MN}\partial_\mu X^M\partial_\nu X^N+
 F_{\mu\nu}+
\partial_\mu T\partial_\nu T \ , 
\mu \ , \nu=0,\dots, p \ ,
\nonumber \\
F_{\mu\nu}=\partial_\mu A_\nu
-\partial_\nu A_\mu \ ,
\nonumber \\
\end{eqnarray}
where $A_\mu \ , 
\mu,\nu=0,\dots,p$ and $ X^{M,N} \ , 
M,N=0,\dots,9$ are gauge and the
transverse scalar
fields on the worldvolume of 
the non-BPS Dp-brane and 
$T$ is the
tachyon field.
$V(T)$ is the tachyon potential that
is symmetric under $T\rightarrow -T$
has maximum at $T=0$ 
equal to the 
tension of a non-BPS Dp-brane 
$\tau_p$
and has its minimum at $T=\pm \infty$
where it vanishes. 

Since we will consider a non-BPS Dp-brane
in the background with nontrivial Ramond-Ramond
field we should also include the 
Wess-Zumino (WZ)
term for non-BPS Dp-brane that is supposed
to have a 
form \cite{Okuyama:2003wm}
\footnote{We work in units $2\pi \alpha'=1$.}
\begin{equation}\label{WZ}
S_{WZ}=\int_{\Sigma}
V(T)\wedge dT\wedge C
e^{F+B} \ .
\end{equation}
In (\ref{WZ}) $\Sigma$ denotes the  worldvolume
of a non-BPS Dp-brane and $C$ collects all
RR n-form gauge potentials (pulled back to the
worldvolume) as
\begin{equation}
C=\oplus_n C_{(n)} \ .
\end{equation}
The form 
of the WZ term (\ref{WZ})
was determined
 from the requirement that
the Ramond-Ramond charge of the tachyon
kink is equal to the charge of
D(p-1)-brane \footnote{Another aspects of 
Wess-Zumino term for non-BPS Dp-brane 
were also discussed in 
\cite{Kraus:2000nj,Kennedy:1999nn,Billo:1999tv,
Takayanagi:2000rz}.}.

Using (\ref{acg}) and (\ref{WZ})
we now  obtain the equations
of motion for $T,X^M$ and $A_\mu$. 
The equation of motion for tachyon
takes the form 
\begin{equation}\label{eqT}
-e^{-\Phi}V'(T)\sqrt{-\det\bA}+
\frac{1}{2}
\partial_\mu 
\left[e^{-\Phi}\partial_\nu T
\left(\bAi^{\nu\mu}+\bAi^{\mu\nu}
\right)\sqrt{-\det\bA}\right]+J_T=0 \ ,
\end{equation}
where $J_T=\frac{\delta}{\delta T} S_{WZ}$ is
the source current derived from varying the Wess-Zumino term.
For scalar modes we obtain
\begin{eqnarray}\label{eqX}
-\frac{\delta e^{-\Phi}}
{\delta X^K}V\sqrt{-\det\bA}-\nonumber \\
-\frac{e^{-\Phi}}{2}
V\left(\frac{\delta g_{MN}}{\delta X^K}
\partial_\mu X^M\partial_\nu X^N+
\frac{\delta b_{MN}}{\delta X^K}
\partial_\mu X^M\partial_\nu X^N\right)
\bAi^{\nu\mu}\sqrt{-\det\bA}+
\nonumber \\
+\frac{1}{2}
\partial_\mu\left[e^{-\Phi}V
g_{KM}\partial_\nu X^M
\left(\bAi^{\nu\mu}+\bAi^{\mu\nu}\right)
\sqrt{-\det\bA}\right]
+\nonumber \\ 
+\frac{1}{2}
\partial_\mu \left[
e^{-\Phi}Vb_{KM}\partial_\nu X^M
\left(\bAi^{\nu\mu}-\bAi^{\mu\nu}\right)
\sqrt{-\det\bA}\right]+J_K=0  \ ,
\nonumber \\
\end{eqnarray}
where $J_K=\frac{\delta}{\delta X^K}
 S_{WZ}$.
Finally, the equations of motion for $A_\mu$ are
\begin{equation}\label{eqA}
\frac{1}{2}
\partial_\nu
\left[e^{-\Phi}V
\left(\bAi^{\mu \nu}-\bAi^{\nu\mu}
\right)\sqrt{-\det\bA}\right]+J^\mu=0 \ ,
\end{equation}
where $J_\mu=\frac{\delta}{\delta A_{\mu}}
 S_{WZ}$.
To  simplify notation it is 
convenient to introduce the
symmetric and antisymmetric form of the 
matrix $\bAi$
\begin{equation}
\bAi_S^{\nu\mu}
=\frac{1}{2}(\bAi^{\nu\mu}+\bAi^{\mu\nu}) \ , \ 
\bAi_A^{\nu\mu}=
\frac{1}{2}(\bAi^{\nu\mu}-\bAi^{\mu\nu}) \ .
\end{equation}
Now we derive the explicit form of the
currents that arise from the WZ term
(\ref{WZ}). To do this 
 we write (\ref{WZ}) as  
(We will closely follow the
analysis of the currents for BPS Dp-brane
that was performed in \cite{Skenderis:2002vf}.)
\begin{equation}\label{swe}
S_{WZ}=\sum_{n\leq 0}
\frac{1}{n!(2!)^nq!}\int
d^{p+1}\xi
\epsilon^{\mu_1\dots \mu_{p+1}}
V(T)\left((\mF)^n_{\mu_1\dots \mu_{2n}}
C_{\mu_{2n+1}\dots \mu_p}
\partial_{\mu_{p+1}}T\right) \ , 
\end{equation}
where $\epsilon^{\mu_1\dots \mu_{p+1}}$ is 
Levi-Civita tensor (with no metric factors)
and $q=(2p+1-1-2n)$. 
The explicit variation of  (\ref{swe})
is equal to 
\begin{eqnarray}
\delta S_{WZ}=
\sum_{n\leq 0}
\frac{1}{n!(2!)^nq!}\int
d^{p+1}\xi
\epsilon^{\mu_1\dots \mu_{p+1}}
\left[V'(T)\delta T\left((\mF)^n_{\mu_1\dots \mu_{2n}}
C_{\mu_{2n+1}\dots \mu_p}
\partial_{\mu_{p+1}}T\right)\right. \nonumber \\
\left.+V(T)\left((\mF)^n_{\mu_1\dots \mu_{2n}}
C_{\mu_{2n+1}\dots \mu_p}
\partial_{\mu_{p+1}}\delta T\right)
+V(T)\left(n(2\partial_{\mu_1}\delta A_{\mu_2}
+\partial_Kb_{MN}\delta X^K\partial_{\mu_1}X^M
\partial_{\mu_2}X^N+\right.\right. 
\nonumber \\
\left.\left.
+2b_{MN}\partial_{\mu_1}\delta X^M
\partial_{\mu_2}X^N)
(\mF)^{n-1}_{\mu_3\dots \mu_{2n}}
C_{\mu_{2n+1}\dots \mu_p}\partial_{\mu_{p+1}}T
\right)
\right. 
\nonumber \\
\left.+
V(T)(\mF)^n_{\mu_1\dots \mu_{2n}}
\left(qC_{M_1\dots M_q}
\partial_{\mu_{2n+1}}\delta X^{M_1}
\dots \partial_{\mu_{p}}X^{M_q}+\right.
\right. \nonumber \\
\left.\left.+
\partial_MC_{M_1\dots M_q}
\delta X^M\partial_{\mu_{2n+1}}X^{M_1}
\dots \partial_{\mu_{p}}X^{M_q}\right)
\partial_{\mu_{p+1}}T\right] \ . 
\nonumber \\
\end{eqnarray}
From this equation we obtain following
form of the currents
\begin{equation}\label{currentA}
J^{\mu_1}=
\sum_{n\geq 0}
\frac{2n}{n! 2^nq!}
\epsilon^{\mu_1\dots \mu_{p+1}}
\partial_{\mu_2}\left[V(T) (\mF)^{n-1}_{\mu_3\dots \mu_{2n}}
C_{\mu_{2n+1}\dots \mu_{p}}\partial_{\mu_{p+1}}T
\right] \ , 
\end{equation}
\begin{eqnarray}\label{currentT}
J_T=\sum_{n\leq 0}
\frac{1}{n!(2!)^nq!}
\epsilon^{\mu_1\dots \mu_{p+1}}
V'(T)\left((\mF)^n_{\mu_1\dots \mu_{2n}}
C_{\mu_{2n+1}\dots \mu_p}
\partial_{\mu_{p+1}}T\right)-\nonumber \\
-\partial_{\mu_{p+1}}
\sum_{n\leq 0}
\frac{1}{n!(2!)^nq!}
\epsilon^{\mu_1\dots \mu_{p+1}}
\left[V(T)(\mF)^n_{\mu_1\dots \mu_{2n}}
C_{\mu_{2n+1}\dots \mu_p}\right] \ .
\nonumber \\ 
\end{eqnarray}
and
\begin{eqnarray}\label{currentXK}
J_K=
\sum_{n\leq 0}
\frac{1}{n!(2!)^nq!}
\epsilon^{\mu_1\dots \mu_{p+1}}
\left[V(T)\partial_Kb_{MN}\partial_{\mu_1}X^M
\partial_{\mu_2}X^N(\mF)^{n-1}_{\mu_3
\dots \mu_{2n}}
C_{\mu_{2n+1}\dots \mu_p}\partial_{\mu_{p+1}}
T\right.
\nonumber \\
\left.+V(T)(\mF)^n_{\mu_1\dots \mu_{2n}}
\partial_K C_{M_1\dots M_q}
\partial_{\mu_{2n+1}}X^{M_1}\dots 
\partial_{\mu_p}X^{M_q}\partial_{\mu_{p+1}}
T-\right.
\nonumber \\
\left.-2n\partial_{\mu_1}
\left[V(T)b_{KM}\partial_{\mu_2}
X^M(\mF)^{n-1}_{\mu_3\dots 
\mu_{2n}}
C_{\mu_{2n+1}\dots \mu_p}\partial_{\mu_{p+1}}T
\right]-\right. \nonumber \\
\left.-q\partial_{2n+1}
\left[V(T)(\mF)^{n}_{\mu_1\dots \mu_{2n}}
C_{KM_2\dots M_q}\partial_{\mu_{2n+2}}
X^{M_2}\dots \partial_{\mu_p}X^{M_q}
\partial_{\mu_{p+1}}T\right]\right]\ . 
\nonumber \\
\end{eqnarray}
Now we try to find 
the solution  of
the  equations 
of motion (\ref{eqT}), (\ref{eqX}) and
(\ref{eqA}) that can 
 be interpreted
as a lower dimensional D(p-1)-brane
 moving in given background. 
Without lost of generality 
we  choose  one
particular worldvolume coordinate,
say $\xi^p\equiv x$
and 
consider following ansatz
for the tachyon
\begin{equation}\label{ansT}
T(x,\xi)=
f(a(x-t(\xi)) \ ,
\end{equation}
where as in \cite{Sen:2003tm} we presume
that $f(u)$ satisfies following
properties
\begin{equation}
f(-u)=-f(u) \ ,
f'(u)>0 \ , \forall u \ ,
f(\pm \infty)=\pm \infty \ 
\end{equation}
but is otherwise an arbitrary 
function of its argument $u$. 
$a$ is a constant that we shall
take to $\infty$ in the end.
In this limit we have $T=\infty$ for
$x>t(\xi)$ and $T=-\infty$ 
for $x<t(\xi)$. 
Note also that $t(\xi)$ in (\ref{ansT})
is function of $\xi^\alpha \ , \alpha=0,\dots,p-1$. 
Let us also   presume following
ansatz for massless fields 
\begin{equation}\label{ansA}
X^M(x,\xi)=X^M(\xi) \ , 
A_x(x,\xi)=0 \ , A_{\alpha}(x,\xi)=
A_\alpha(\xi) \ ,
\alpha=0,\dots,p-1 \ ,  
\end{equation}
where again $\xi\equiv (\xi^0,\dots,\xi^{p-1})$.
Before we proceed further we would
like to stress what is the main goal of
this analysis. We would like to
show that the  dynamics 
of the kink 
is governed by the action
\begin{eqnarray}\label{BPSact}
S=S_{DBI}+S_{WZ}  \ , \nonumber \\
S_{DBI}=-T_{p-1}\int d^p\xi
e^{-\Phi}\sqrt{-\det\mat} \ , 
\nonumber \\
S_{WZ}=\sum_{n\leq 0}
\frac{1}{n!(2!)^nq!}\int
d^p\xi
\epsilon^{\alpha_1\dots \alpha_p}
(\mtF)^n_{\alpha_1\dots \alpha_{2n}}
\mtC_{\alpha_{2n+1}\dots \alpha_p}
 \ , \nonumber \\ 
\end{eqnarray}
where
\begin{eqnarray}
\mat_{\alpha\beta}=
(g_{MN}+b_{MN})\partial_\alpha X^M\partial_\beta X^N+
F_{\alpha\beta} \ , \nonumber \\
\mtF_{\alpha\beta}=F_{\alpha\beta}+
b_{MN}\partial_\alpha X^M\partial_\beta
X^N \  \ , \nonumber \\
\mtC_{\alpha_{2n+1}\dots \alpha_p}=
C_{M_{2n+1}\dots M_p}\partial_{\alpha_{2n+1}}
X^{M_{2n+1}}\dots\partial_{\alpha_p}X^{M_p} \ .
\nonumber \\ 
\end{eqnarray}
In other words we will show that
the  modes given in
 (\ref{ansA}) that propagate on
the worldvolume of the kink obey 
the  equations of motion derived
from (\ref{BPSact}) that 
have the form
\begin{eqnarray}\label{eqXbps}
-\frac{\delta e^{-\Phi}}
{\delta X^K}V\sqrt{-\det\bA}-\nonumber \\
-\frac{e^{-\Phi}}{2}
\left(\frac{\delta g_{MN}}{\delta X^K}
\partial_\alpha X^M\partial_\beta X^N+
\frac{\delta b_{MN}}{\delta X^K}
\partial_\alpha X^M\partial_\beta X^N\right)
\mati^{\beta\alpha}\sqrt{-\det\mat}+
\nonumber \\
+\partial_\alpha\left[e^{-\Phi}
g_{KM}\partial_\beta X^M
\mati_S^{\beta\alpha}
\sqrt{-\det\mat}\right]
+\nonumber \\ 
+\partial_\alpha \left[
e^{-\Phi}b_{KM}\partial_\beta X^M
\mati^{\beta\alpha}_A
\sqrt{-\det\mat}\right]+\tilde{J}_K=0  \ ,
\nonumber \\
\end{eqnarray}
where 
\begin{eqnarray}\label{currentxkbps}
\tilde{J}_K=\frac{\delta S_{WZ}}
{\delta X^K}=
\sum_{n\leq 0}
\frac{1}{n!(2!)^nq!}
\epsilon^{\alpha_1\dots \alpha_p}
\left[\partial_Kb_{MN}\partial_{\alpha_1}X^M
\partial_{\alpha_2}X^N(\mtF)^{n-1}_{\alpha_3
\dots \alpha_{2n}}
\mtC_{\alpha_{2n+1}\dots \alpha_p}\right.
\nonumber \\
\left.+(\mtF)^n_{\alpha_1\dots \alpha_{2n}}
\partial_K \mtC_{M_1\dots M_q}
\partial_{\alpha_{2n+1}}X^{M_1}\dots 
\partial_{\alpha_p}X^{M_q}-\right.
\nonumber \\
\left.-2n\partial_{\alpha_1}
\left[b_{KM}\partial_{\alpha_2}
X^M(\mtF)^{n-1}_{\alpha_3\dots 
\alpha_{2n}}
\mtC_{\alpha_{2n+1}\dots \alpha_p}
\right]-\right. \nonumber \\
\left.-q\partial_{\alpha_{2n+1}}
\left[(\mtF)^{n}_{\alpha_1\dots \alpha_{2n}}
C_{KM_2\dots M_q}\partial_{\alpha_{2n+2}}
X^{M_2}\dots \partial_{\alpha_p}X^{M_q}
\right]\right]\ . 
\nonumber \\
\end{eqnarray}
In the same way we get
that the equation of
 motion for $A_\alpha$ are
\begin{equation}\label{eqAbps}
\partial_\beta
\left[e^{-\Phi}
\mati^{\alpha\beta}_A
\sqrt{-\det\mat}\right]+\tilde{J}^\alpha=0 \ ,
\end{equation}
where
\begin{equation}\label{currentAbps}
\tilde{J}^{\alpha_1}=
\sum_{n\geq 0}
\frac{2n}{n! 2^nq!}
\epsilon^{\alpha_1\dots\alpha_p}
\partial_{\alpha_2}\left[ (\mtF)^{n-1}_{\alpha_3
\dots \alpha_{2n}}
\mtC_{\alpha_{2n+2}\dots \alpha_p}
\right] \  .
\end{equation}
Let us again return to the
ansatz (\ref{ansT}) and (\ref{ansA})
and calculate the matrix  $\bA_{\mu\nu}$
\begin{equation}\label{bA}
\bA=\left(\begin{array}{cc}
\mat_{\alpha\beta}+
a^2f'^2\partial_\alpha t
\partial_\beta t & -a^2f'^2\partial_\beta t \\
-a^2f'^2\partial_\alpha t & a^2f'^2 \\
\end{array}\right)
\end{equation}
where 
\begin{equation}
\mat_{\alpha\beta}=
(g_{MN}+b_{MN})
\partial_\alpha X^M\partial_\beta X^N+
F_{\alpha\beta} \ . 
\end{equation}
Now using the fact that
\begin{equation}
\det\bA=
\det(\bA_{\alpha\beta}-
\bA_{\alpha x}\frac{1}{\bA_{xx}}
\bA_{x\beta})\det\bA_{xx}
\end{equation}
we get
\begin{equation}
\det\bA=a^2f'^2\det\mat \ .
\end{equation}
As a next step we  determine
the inverse matrix $\bAi$ up
the correction $\mathcal{O}\left(\frac{1}{a}
\right)$. After some algebra
we get
\begin{eqnarray}\label{bai}
\bAi^{\alpha\beta}=\mati^{\alpha\beta} \ ,
\bAi^{x\beta}=\partial_\alpha t
\mati^{\alpha\beta} \ , \nonumber \\
\bAi^{\alpha x}=\mati^{\alpha \beta}
\partial_\beta t \ ,
\bAi^{xx}=\partial_\alpha t\mati^{\alpha\beta}
\partial_\beta t \  \nonumber \\
\end{eqnarray}
In the limit of large $a$. Using
also the relation $\bA_{\mu\nu}\bAi^{\nu\rho}=
\delta_\mu^{\ \rho}$ and the form
of the matrix $\bA$ given in  (\ref{bA})
we easily determine 
following relation
\begin{equation}\label{baip}
\bAi^{\mu x}_S-\bAi^{\mu\alpha}_S
\partial_\alpha t=\frac{1}{a^2f'^2}
\left(\delta^{\mu}_x-\bAi^{x\mu}_S\right) \ . 
\end{equation}
Now with the
help of (\ref{baip}) we get
\begin{eqnarray}\label{dbith}
\partial_\mu\left[e^{-\Phi}V
\partial_\nu T\bAi^{\nu\mu}_S
\sqrt{-\det\bA}\right]=
\nonumber \\
\partial_\mu \left[
e^{-\Phi}Vaf'(\bAi^{x\mu}_S-
\bAi^{\alpha\mu}_S
\partial_\alpha t)\sqrt{-\det\bA}\right]=
\nonumber \\
=\partial_\mu \left[
e^{-\Phi}V(\delta^{\mu}_x-
\bAi^{x\mu}_S)\sqrt{-\det\mat}\right]=
\nonumber \\
\partial_x\left[e^{-\Phi}
V(T)(1-\bAi^{xx}_S)\sqrt{-\det\mat}\right]-
\partial_\alpha\left[e^{-\Phi}
V\bAi^{x\alpha}_S
\sqrt{-\det\mat}\right]
=\nonumber \\
=V'af'e^{-\Phi}
(1-\mati^{\alpha\beta}_S
\partial_\alpha t\partial_\beta t)
\sqrt{-\det\mat}
+V'af'\partial_\alpha t\mati^{\alpha\beta}_S
\partial_\beta t \sqrt{-\det\mat}
-\nonumber \\
-V\partial_\alpha
\left[e^{-\Phi}\mati^{\beta\alpha}_S
\partial_\beta t\sqrt{-\det\mat}\right]=
\nonumber \\
=V'af'e^{-\Phi}\sqrt{-\det\mat}
-V\partial_\alpha
\left[e^{-\Phi}\mati^{\beta\alpha}_S
\partial_\beta t\sqrt{-\det\mat}\right] \ ,
\nonumber \\
\end{eqnarray}
where we have used the fact
that $\partial_\alpha V=V'\partial_\alpha T=
-V'f'\partial_\alpha t$ and also the
fact that the only field that depends on $x$ is
tachyon $T$. 
Using (\ref{dbith}) we get following
form of the  DBI part of the tachyon
 equation of motion (\ref{eqT})
\begin{equation}\label{dbitf}
V\partial_\alpha\left[
e^{-\Phi}\mati^{\beta\alpha}_S
\partial_\beta t\sqrt{-\det\mat}\right] \ . 
\end{equation}
Now we consider the DBI part of the
equation of motion for 
$X^K$ (\ref{eqX}). 
With the ansatz (\ref{ansT})
and (\ref{ansA}) 
the first two lines there take the form
\begin{eqnarray}
-af'V\partial_K [e^{-\Phi}]\sqrt{-\det\mat}
\nonumber \\
--af'V\frac{e^{-\Phi}}{2}
\left(\partial_Kg_{MN}+\partial_Kb_{MN}\right)
\partial_\alpha X^M\partial_\beta X^N\mati^{\beta\alpha}
\sqrt{-\det\mat} \ .  \nonumber \\
\end{eqnarray}
On the other hand the expression on the
third line in (\ref{eqX}) takes the form
\begin{eqnarray}\label{htx}
\partial_\mu\left[e^{-\Phi}Vg_{KM}\partial_\alpha X^M
\bAi^{\alpha \mu}_Saf'\sqrt{-\det\mat}\right]=
\nonumber \\
=aVf'\partial_\beta\left[
e^{-\Phi}g_{KM}\partial_\alpha
X^M \mati^{\alpha\beta}_S 
\sqrt{-\det\mat}\right] \ , 
\nonumber \\
\end{eqnarray}
where we have used 
\begin{eqnarray}\label{pdvf}
\partial_\beta[aVf']=
-\partial_x[aVf']\partial_\beta t \ 
\nonumber \\
\end{eqnarray}
and also the fact that $X^K$ are
function of $\xi^\alpha$ only. 
In the same way as in  (\ref{htx})
we can show that
\begin{eqnarray}\label{htx1}
\partial_\mu\left[e^{-\Phi}Vb_{KM}\partial_\alpha X^M
\bAi^{\alpha \mu}_Aaf'\sqrt{-\det\mat}\right]=
\nonumber \\
=af'V
\partial_\beta\left[e^{-\Phi}
b_{KM}\partial_\alpha X^M\mati^{\beta\alpha}_A
\sqrt{-\det\mat}\right] \ . 
\nonumber \\
\end{eqnarray}
If we collect all these results
 we obtain
that the DBI part of the equation of motion for $X^K$ takes
the form
\begin{eqnarray}\label{dbixf}
af'V
\left(-\partial_K[e^{-\Phi}]
\sqrt{-\det\mat}-\frac{e^{-\Phi}}{2}
\left(\partial_Kg_{MN}+\partial_Kb_{MN}\right)
\partial_\alpha X^M\partial_\beta X^N\mati^{\beta\alpha}
\sqrt{-\det\mat}\right. \nonumber \\
\left. 
+\partial_\beta\left[
e^{-\Phi}g_{KM}\partial_\alpha
X^M \mati^{\alpha\beta}_S
\sqrt{-\det\mat}\right]+
\partial_\beta\left[e^{-\Phi}
b_{KM}\partial_\alpha X^M\mati^{\alpha\beta}_A
\sqrt{-\det\mat}\right]\right) \ . 
\nonumber \\
\end{eqnarray}
Now let us consider the equation of motion for
gauge field. For $A_x$ we get
\begin{eqnarray}\label{eqax}
\partial_\nu\left[Ve^{-\Phi}\bAi
^{x\nu}_A\sqrt{-\det\bA}\right]=
af'V\partial_\beta t \partial_\alpha
\left[e^{-\Phi}\mati^{\beta\alpha}_A
\sqrt{-\det\mat}\right] \ , 
\nonumber \\
\end{eqnarray}
where we have used  an 
 antisymmetry of $\mati^{\alpha\beta}_A$ so that
$\mati^{\alpha\beta}_A\partial_\alpha \partial_\beta t=0$.

On the other hand the equations of motion for $A_\alpha$ 
take the form
\begin{eqnarray}\label{dbiaf}
\partial_\mu\left[e^{-\Phi}
\bAi^{\alpha \mu}_A\sqrt{-\det\bAi}\right]=
\partial_x[af'V]e^{-\Phi}\mati^{\alpha\beta}_A\partial_\beta t
\sqrt{-\det\mat}+\nonumber \\
\partial_\beta\left[af'V
e^{-\Phi}\mati^{\alpha\beta}_A\sqrt{-\det\mat}\right]=
af'V\partial_\beta\left[
e^{-\Phi}\mati^{\alpha\beta}_A\sqrt{-\det\mat}\right] \ .
\nonumber \\
\end{eqnarray}
As a next step we evaluate the
currents given in (\ref{currentA}), (\ref{currentT})
and (\ref{currentXK}) for the 
ansatz (\ref{ansT}) and (\ref{ansA}). 

To begin with  we determine the components
of the embedding of various fields.
It is easy to see that
\begin{equation}
\mF_{x\alpha}=-\mF_{\alpha x}=
F_{x\alpha}+b_{MN}\partial_xX^M\partial_\alpha X^N=0
\end{equation}
due to the fact that all worldvolume massless
 modes do not depend on $x$ and
also thanks to the fact that $A_x=0$. Then the only
nonzero components of $\mF_{\mu\nu}$
 are $\mtF_{\alpha\beta}$. 
For $C^{(n)}$ the situation is the same, namely any component
that in the subscript contains
 $x$ vanishes
since
\begin{equation}
C_{\dots x\dots}=
C_{\dots M\dots}\partial_x X^M=0 \ .
\end{equation}

Now we begin with the gauge current $J^{\mu}$. Firstly,
$J^x$ is equal to
\begin{eqnarray}\label{currentax}
J^x=\sum_{n\geq 0}
\frac{2n}{n! 2^nq!}
\epsilon^{x\alpha_2\alpha_3\dots \alpha_p\alpha_1}
\partial_{\alpha_2}\left[V(T) (\mtF)^{n-1}_{\alpha_3
\dots \alpha_{2n}}
\mtC_{\alpha_{2n+1}\dots \alpha_{p}}\partial_{\alpha_1}T
\right]=
\nonumber \\
-\sum_{n\geq 0}
\frac{2n}{n! 2^nq!}
\epsilon^{x\alpha_2\alpha_3\dots \alpha_p\alpha_1}
\partial_{\alpha_1}[Vaf']\left[(\mtF)^{n-1}_{\alpha_3
\dots \alpha_{2n}}
\mtC_{\alpha_{2n+1}\dots \alpha_{p}}\partial_{\alpha_1}t
\right]-\nonumber \\
-af'V\sum_{n\geq 0}
\frac{2n}{n! 2^nq!}
\epsilon^{x\alpha_2\alpha_3\dots \alpha_p\alpha_1}
\partial_{\alpha_2}\left[(\mtF)^{n-1}_{\alpha_3
\dots \alpha_{2n}}
\mtC_{\alpha_{2n+1}\dots \alpha_{p}}\partial_{\alpha_1}t\right]
\nonumber \\
=af'V\partial_{\alpha_1}t\sum_{n\geq 0}
\frac{2n}{n! 2^nq!}
\epsilon^{\alpha_1\dots \alpha_px}
\partial_{\alpha_2}\left[(\mtF)^{n-1}_{\alpha_3
\dots \alpha_{2n}}
\mtC_{\alpha_{2n+1}\dots \alpha_{p}}\right]=
af'V\partial_{\alpha_1}t\tilde{J}^{\alpha_1}
 \ , 
\nonumber \\ 
\end{eqnarray}
where we have used the fact that $\partial_{\alpha_1}(Vf')=
-\partial_x(Vf')\partial_{\alpha_1}t$ and then 
an antisymmetry of $\epsilon^{x\alpha_1\dots\alpha_p}$
so that $\epsilon^{\alpha_1\alpha_2\dots}\partial_{\alpha_1}
\partial_{\alpha_2}t=0$. 
Also the form of the current   $\tilde{J}^{\alpha}$
was  given in (\ref{currentAbps})
\footnote{Note that $\epsilon^{\alpha_1\dots
\alpha_p x}=\epsilon^{\alpha_1\dots\alpha_p}$.}.
 
On the other hand the current $J^{\alpha_1}$ is
equal to
\begin{eqnarray}\label{currentAf}
J^{\alpha_1}=\sum_{n\geq 0}
\frac{2n}{n! 2^nq!}
\epsilon^{\alpha_1\alpha_2\dots \alpha_p x}
\partial_{\alpha_2}\left[V(T) (\mtF)^{n-1}_{\alpha_3\dots \alpha_{2n}}
\mtC_{\alpha_{2n+1}\dots \alpha_{p}}\partial_{x}T
\right]+
\nonumber \\
+\sum_{n\geq 0}\frac{2n}{n! 2^nq!}
\epsilon^{\alpha_1x\alpha_3\dots \alpha_p \alpha_2}
\partial_{x}\left[V(T) (\mtF)^{n-1}_{\alpha_3\dots \alpha_{2n}}
\mtC_{\alpha_{2n+1}\dots \alpha_{p}}\partial_{\alpha_2}T
\right]=\nonumber \\
=\sum_{n\geq 0} af'V\frac{2n}{n! 2^nq!}
\epsilon^{\alpha_1\alpha_2\dots \alpha_p x}
\partial_{\alpha_2}\left[(\mtF)^{n-1}_{\alpha_3\dots \alpha_{2n}}
\mtC_{\alpha_{2n+1}\dots \alpha_{p}}\right]=
af'V\tilde{J}^{\alpha_1} \nonumber \\
\end{eqnarray}
using the fact that $(\mF)^{n-1}_{\dots x}$
and $C_{\dots x\dots}$ 
are equal to zero. 
  If we now combine (\ref{dbiaf})
with (\ref{currentAf})  we get
\begin{equation}\label{eqaf}
af'V
\left[\partial_\beta\left[
e^{-\Phi}\mati^{\alpha\beta}_A\sqrt{-\det\mat}\right]+
\tilde{J}^{\alpha}\right]=0 \ . 
\end{equation}
Let us now analyse the behaviour of the
term  $af'V$ in the
limit $a\rightarrow \infty$. Since by definition
$f'(u)$ is finite for all 
$u$ it remains
to study the properties of the expression
$aV$. Since $V\sim e^{-T}$ for $T\rightarrow 
\infty$ we have
\begin{eqnarray}
\lim_{a\rightarrow \infty} aV(f(a(x-t(\xi))=
 \ (\mathrm{for} \ x\neq t(\xi))
 \nonumber \\
 \lim_{a\rightarrow \infty} 
 \frac{a}{e^{f(a(x-t(\xi)))}}=
\frac{1}{(x-t(\xi))f'}\lim_{a\rightarrow \infty}
e^{-f(a(x-t(\xi)))}=0 \ . \nonumber \\
\end{eqnarray}
We see that for $x\neq t(\xi)$ the
expression $aV$ goes to zero in
the limit $a\rightarrow \infty$. On the
other hand for $x=t(\xi)$ the potential
$V(0)=\tau_p$ and hence   in order to obey the
equation of motion for $A_{\alpha}$ we
find that the expression in the bracket in 
(\ref{eqaf}) should vanish.
 In fact, this
expression is the same as the equation of motion
for $A_{\alpha}$ given in (\ref{eqAbps})

On the other hand using 
(\ref{eqax}) and  (\ref{currentax}) 
 the equation of motion for $A_x$ takes the form
\begin{equation}
af'V
\left[\partial_\beta t \left(\partial_\alpha
\left[e^{-\Phi}\mati^{\beta\alpha}_A
\sqrt{-\det\mat}\right]+\tilde{J}^\beta\right)
\right]=0 \  
\end{equation} 
that  clearly holds using the fact that 
all modes obey the equation of motion (\ref{eqAbps}).

Now we will analyse the current $J_K$. Looking
on its form  (\ref{currentXK})
it is clear that the expressions
on the first and the second line are nonzero for
$\mu_{p+1}=x$ only. On the other hand the 
expression on the third line can be nonzero for
$\mu_{p+1}=x$ and for $\mu_1=x$ where we
get
\begin{eqnarray}\label{xknf}
-2n\epsilon^{\alpha_1
\dots \alpha_p x}\partial_{\alpha_1}
\left[aVf'b_{KM}\partial_{\alpha_2}
X^M(\mtF)^{n-1}_{\alpha_3\dots 
\alpha_{2n}}
\mtC_{\alpha_{2n+1}\dots \alpha_p}
\right]+\nonumber \\
+2n\epsilon^{x\alpha_2
\dots \alpha_p \alpha_1}\partial_{x}
[aVf']\left[\partial_{\alpha_1}t 
b_{KM}\partial_{\alpha_2}
X^M(\mtF)^{n-1}_{\alpha_3\dots 
\alpha_{2n}}
\mtC_{\alpha_{2n+1}\dots \alpha_p}
\right]=\nonumber \\
=-af'V2n\epsilon^{\alpha_1
\dots \alpha_p x}\partial_{\alpha_1}
\left[b_{KM}\partial_{\alpha_2}
X^M(\mtF)^{n-1}_{\alpha_3\dots 
\alpha_{2n}}
\mtC_{\alpha_{2n+1}\dots \alpha_p}
\right] \ .  \nonumber \\
\end{eqnarray}
Finally, the  expression on the last
line in (\ref{currentXK}) is equal to
\begin{eqnarray}\label{xkqf}
-q\epsilon^{\mu_1
\dots\mu_{p+1}}\partial_{\mu_{2n+1}}
\left[V(T)(\mF)^{n}_{\mu_1\dots \mu_{2n}}
C_{KM_2\dots M_q}\partial_{\mu_{2n+2}}
X^{M_2}\dots \partial_{\mu_p}X^{M_q}
\partial_{\mu_{p+1}}T\right]=
\nonumber \\
=-q\epsilon^{\alpha_1
\dots\alpha_px}\partial_{\alpha_{2n+1}}
\left[af'V(T)(\mtF)^{n}_{\alpha_1\dots \alpha_{2n}}
C_{KM_2\dots M_q}\partial_{\alpha_{2n+2}}
X^{M_2}\dots \partial_{\alpha_p}X^{M_q}
\right]+\nonumber \\
q\epsilon^{\alpha_1
\dots \alpha_{2n}
x\alpha_{2n+2}\dots \alpha_p\alpha_{2n+1}}\partial_{x}
\left[V(T)af'\partial_{\alpha_{2n+1}}t
(\mtF)^{n}_{\mu_1\dots \mu_{2n}}
C_{KM_2\dots M_q}\partial_{\mu_{2n+2}}
X^{M_2}\dots \partial_{\mu_p}X^{M_q}
\right]=\nonumber \\
=-af'Vq\epsilon^{\alpha_1
\dots\alpha_px}\partial_{\alpha_{2n+1}}
\left[(\mtF)^{n}_{\alpha_1\dots \alpha_{2n}}
C_{KM_2\dots M_q}\partial_{\alpha_{2n+2}}
X^{M_2}\dots \partial_{\alpha_p}X^{M_q}
\right] \ .  \nonumber \\ 
\end{eqnarray}
If we now combine all these results
together  we obtain
 final form of the current $J_K$
\begin{eqnarray}\label{currentXKf}
J_K=af'V\sum_{n\leq 0}
\frac{1}{n!(2!)^nq!}
\epsilon^{\alpha_1\dots \alpha_px}
\left(\partial_Kb_{MN}\partial_{\alpha_1}X^M
\partial_{\alpha_2}X^N(\mtF)^{n-1}_{\alpha_3
\dots \alpha_{2n}}
\mtC_{\alpha_{2n+1}\dots \alpha_p}
\right.
\nonumber \\
\left.+
(\mtF)^n_{\alpha_1\dots \alpha_{2n}}
\partial_K C_{M_1\dots M_q}
\partial_{\alpha_{2n+1}}X^{M_1}\dots 
\partial_{\alpha_p}X^{M_q}-\right.
\nonumber \\
\left.-2n\partial_{\alpha_1}
\left[b_{KM}\partial_{\alpha_2}
X^M(\mtF)^{n-1}_{\alpha_3\dots 
\alpha_{2n}}
\mtC_{\alpha_{2n+1}\dots \alpha_p}
\right]+\right.
 \nonumber \\
\left.+q\partial_{\alpha_{2n+1}}
\left[(\mtF)^{n}_{\alpha_1\dots \alpha_{2n}}
C_{KM_2\dots M_q}\partial_{\alpha_{2n+2}}
X^{M_2}\dots \partial_{\alpha_p}X^{M_q}
\right]\right)\equiv af'V\tilde{J}_K \ ,  \nonumber \\ 
\end{eqnarray}
where $\tilde{J}_K$ was defined in 
(\ref{currentxkbps}).
Using (\ref{dbixf}) and (\ref{currentXKf})
we obtain the final form of the
 equation of motion for $X^K$
in  the form
\begin{eqnarray}\label{eqXKf}
af'V
\left(-\partial_K[e^{-\Phi}]
\sqrt{-\det\mat}-\right.
\nonumber \\
\left.-\frac{e^{-\Phi}}{2}
\left(\partial_Kg_{MN}+\partial_Kb_{MN}\right)
\partial_\alpha X^M\partial_\beta X^N\mati^{\alpha\beta}
\sqrt{-\det\mat}\right. \nonumber \\
\left. 
+\partial_\beta\left[
e^{-\Phi}g_{KM}\partial_\alpha
X^M \mati^{\alpha\beta}_S
\sqrt{-\det\mat}\right]+\right. \nonumber\\
\left.+
\partial_\beta\left[e^{-\Phi}
b_{KM}\partial_\alpha X^M\mati^{\beta\alpha}_A
\sqrt{-\det\mat}\right]
+\tilde{J}_K\right)=0 \ . 
\nonumber \\
\end{eqnarray}
Following  discussion given 
below (\ref{eqaf}) we see that
the expression in the bracket in 
(\ref{eqXKf}) should be equal to zero. 
On the other hand this equation is exactly
the equation of motion for the embedding
mode that lives on the worldvolume
of D(p-1)-brane that was
 given in (\ref{eqXbps}).

Finally we come to the analysis of the
tachyon current $J_T$ that can be written as
\begin{eqnarray}
J_T=-\sum_{n\leq 0}V(T)\frac{1}{n!(2!)^nq!}
\epsilon^{\mu_1\dots \mu_{p+1}}
\partial_{\mu_{p+1}}\left((\mF)^n_{\mu_1\dots \mu_{2n}}
C_{\mu_{2n+1}\dots \mu_p}
\right) \ . 
\nonumber \\
\end{eqnarray}
It is not hard to see that the
 tachyon current is equal to zero. Firstly,
the contribution to $J^T$ for which  $\mu^{p+1}=x$ 
vanishes thanks to the
fact that all massless modes
do not depend on $x$. On the
other hand for $\mu_{p+1}\neq x$ all contributions
to $J_T$
vanish  since then there  certainly exists $\mF$ or
$C$ with the lower index containing $x$ and as
we argued above these terms are equal to zero.
Hence we get
\begin{equation}\label{currentTf}
J_T=0 \ .
\end{equation}
Then the equation (\ref{eqT}) takes the form
\begin{equation}
V\partial_\alpha\left[
e^{-\Phi}\mati^{\beta\alpha}_S
\partial_\beta t\sqrt{-\det\mat}\right]=0 \ . 
\end{equation}
Since for general background all massless fields 
depend on $\xi$ the only way how
to obey this equation for $x=t(\xi)$ where
$V(0)=\tau_p$ is to demand that 
$\partial_\alpha t=0$. 
In other words  we obtain a set
 of the tachyon kink solutions
labelled  with constant $t$ that determines
the position of the
 core of the kink on the worldvolume of
an unstable Dp-brane.  We 
mean that this is a natural result 
for a non-BPS Dp-brane where
no gauge fixing procedure was  imposed.
In  this case the  position  of  a Dp-brane
in the target spacetime  is  not specified 
and consequently all kink solutions on its
worldvolume  are equivalent. 

In summary,   we have shown that 
the spatial dependent tachyon
condensation on the worldvolume of
a unstable Dp-brane in general
background  leads to 
an emergence of a lower dimensional D(p-1)-brane 
 where the massless
modes that propagate on the worldvolume
of the kink obey the equations of motion
that arise from the DBI and WZ action for
D(p-1)-brane.  
\subsection{Stress energy tensor}
Further support for an 
interpretation
of the tachyon kink as 
a lower dimensional
D(p-1)-brane 
can be derived  
 from 
the analysis
of the stress energy tensor
for the non-BPS Dp-brane.
In order to find its form  recall 
that   we 
can write the action (\ref{acg})
as 
\begin{equation}\label{dactem}
S_{p}=-\int d^{10}xd^{(p+1)}
\xi\delta
(X^M(\xi)-x^M)e^{-\Phi}V(T)
\sqrt{-\det \bA} \ .
\end{equation}
From  (\ref{dactem})
we can easily determine components
of the stress energy 
tensor $T_{MN}(x)$ of an unstable
D-brane using the fact that the stress
energy tensor $T_{MN}(x)$  is defined 
as the variation of $S_p$
with   
respect
to $g_{MN}(x)$ 
\begin{eqnarray}\label{TMNg}
T_{MN}(x)=-2
\frac{\delta S_{p}}{
\sqrt{-g(x)}\delta g^{MN}(x)}=\nonumber \\
=-\int d^{(p+1)}\xi\frac{\delta(X^M(\xi)
-x^M)}
{\sqrt{-g(x)}}e^{-\Phi}V
g_{MK}g_{NL}
\partial_{\mu}X^K\partial_{\nu}
X^L(\bA^{-1})^{\nu\mu}_S
\sqrt{-\det \bA} \ . \nonumber \\ 
\end{eqnarray}
Now from (\ref{ansT}) and (\ref{ansA})
we  know that all massless modes are $x$ independent.
Hence (\ref{TMNg}) is equal to
\begin{eqnarray}
T_{MN}(x)=-\int dx af'V(f(x))
\int d^{p}\xi\frac{\delta(X^M(\xi)
-x^M)}
{\sqrt{-g(x)}}\times \nonumber \\
\times e^{-\Phi}
g_{MK}g_{NL}
\partial_{\alpha}X^K\partial_{\beta}
X^L\mati^{\beta\alpha}_S
\sqrt{-\det \mat}= \nonumber \\
-T_{p-1} \int d^{p}\xi\frac{\delta(X^M(\xi)
-x^M)}
{\sqrt{-g(x)}}e^{-\Phi}
g_{MK}g_{NL}
\partial_{\alpha}X^K\partial_{\beta}
X^L\mati^{\beta\alpha}_S
\sqrt{-\det \mat} \ ,    \nonumber \\
\end{eqnarray}
where
\begin{equation}
T_{p-1}=\int dx aV(f)f'=
\int dm V(m)
\end{equation}
is a tension of BPS D(p-1)-brane. 
In other words the stress energy tensor
evaluated on the ansatz (\ref{ansT}) and
(\ref{ansA}) corresponds to the
stress energy tensor
 for D(p-1)-brane. 

In the same way we can
study 
 other currents that express the coupling
of the non-BPS Dp-brane to  closed
string massless fields. For example, let us 
consider current $J^{M_1\dots M_N}_C$ corresponding
to the variation of $S_{WZ}$ with respect  to
$C_{M_1\dots M_N}(x)$
\begin{eqnarray}
J^{M_1\dots M_N}_C(x)=
\frac{1}{n! (2!)^nN!}\int d^{p+1}\xi
\delta^{10}(x^M-X^M(\xi))
V(T)\epsilon^{\mu_1\dots \mu_{p+1}}\times
\nonumber \\
\times
(\mF)^n_{\mu_1\dots \mu_{2n}}
\partial_{\mu_{2n+1}}X^{M_1}\dots \partial_{\mu_p}
X^{M_N}\partial_{\mu_{p+1}}T \ , \nonumber \\
\end{eqnarray}
where $n=\frac{p-N}{2}$.
It is clear that the nonzero components corresponds
to $\mu_{p+1}=x$ (since in the opposite case
there will be derivative $\partial_x X$ that for
(\ref{ansA}) vanishes)
and we get
\begin{eqnarray}\label{jmn}
J^{M_1\dots M_N}_C(x)=
\int dx V(f)f'a\frac{1}{n! (2!)^nN!}\int d^p\xi
\delta^{10}(x^M-X^M(\xi))
 \epsilon^{\alpha_1\dots \alpha_p x}\times
\nonumber \\
\times
(\mF)^n_{\alpha_1\dots \alpha_{2n}}
\partial_{\alpha_{2n+1}}
X^{M_1}\dots \partial_{\alpha_p}
X^{M_N} \nonumber \\
=\frac{\mu_{p-1}}{n! (2!)^nN!}\int d^p\xi
\delta^{10}(x^M-X^M(\xi))
\epsilon^{\alpha_1\dots \alpha_p x}\times
\nonumber \\
\times
(\mF)^n_{\alpha_1\dots \alpha_{2n}}
\partial_{\alpha_{2n+1}}
X^{M_1}\dots \partial_{\alpha_p}
X^{M_N} \ , \nonumber \\
\end{eqnarray}
where $\mu_{p-1}=T_{p-1}$ is a 
Ramond-Ramond charge of
D(p-1)-brane and hence  (\ref{jmn})
is an appropriate current for D(p-1)-brane. 
\section{Partial fixing gauge}\label{third}
 In order
to find   solution of the tachyon effective
action, where
the  mode $t$ that determines the location
of the core of the kink  could be 
interpreted as  an additional embedding
coordinate, we  should  
partial fix the gauge. 
In other words,  when we choose one
spatial coordinate on the worldvolume theory
 on which the tachyon depends we will
also  presume that this coordinate
coincides with one arbitrary spatial coordinate
in the target spacetime.
Since both worldvolume theory and 
spacetime theory are diffeomorphism invariant
we can without loose of generality choose
the worldvolume direction on which the tachyon
depends to be $\xi^p$ and 
the spacetime direction to be $X^9$.
Then we
demand that
\begin{equation}
X^9=x\equiv \xi^p \ . 
\end{equation}
Let us now 
consider following ansatz
for the tachyon
\begin{equation}\label{ansTf}
T(x,\xi)=
f(a(x-t(\xi)) \ ,
\end{equation}
where $f(u)$ could be the
same 
function as was defined in
previous section. We also
 presume following ansatz
for massless modes
\begin{equation}\label{ansAf}
X^I(x,\xi)=X^I(\xi) \ , 
A_x(x,\xi)=0 \ , A_{\alpha}(x,\xi)=
A_\alpha(\xi)  \ , 
\end{equation}
where $I,J,K=0,1\dots, 8$
and where 
 $\xi^\alpha \ ,
\alpha=0,\dots,p-1$ are coordinates
tangential to the kink worldvolume.

Now we must show that the 
ansatz (\ref{ansTf}) and (\ref{ansAf}) solve the
equation of motion 
for $T, X^M$ and $A_\mu$. 
Firstly, for (\ref{ansTf}) 
and (\ref{ansAf}) the
matrix 
$\bA_{\mu\nu}$ takes the form
\begin{eqnarray}\label{bAf}
\bA_{xx}=g_{99}+a^2f'^2 \ , 
\bA_{x\beta}=g_{9I}\partial_\beta X^I
+b_{9I}\partial_\beta X^I
-a^2f'^2\partial_\beta t\equiv 
H_{x\beta}-a^2f'^2\partial_\beta t 
 \ , \nonumber \\
\bA_{\alpha x}=
\partial_\alpha X^Ig_{I9}+
\partial_\alpha X^Ib_{I9}
-a^2f'^2\partial_\alpha t\equiv
H_{\alpha x}-a^2f'^2\partial_\alpha t
 \ ,
\nonumber \\
\bA_{\alpha\beta}= 
(a^2f'^2-g_{99})\partial_\alpha t\partial_\beta t
-H_{\alpha x}\partial_\beta t-
\partial_\alpha tH_{x\beta}
+\mat_{\alpha\beta}
\ , \nonumber \\
\mat_{\alpha \beta}=
g_{99}\partial_\alpha t\partial_\beta t+
g_{IJ}\partial_\alpha X^I\partial_\beta X^J
+\partial_\alpha X^Ig_{I9}\partial_\beta t
+\partial_\alpha t g_{9J}\partial_\beta X^J+
\nonumber \\
+b_{IJ}\partial_\alpha X^I\partial_\beta X^J+
\partial_\alpha X^Ib_{I9}\partial_\beta t
+\partial_\alpha t b_{9J}\partial_\beta X^J
+F_{\alpha \beta} \ . \nonumber \\
\end{eqnarray}
As in the previous section we obtain
that $\det\bA$ is equal to 
\begin{equation}\label{dea}
\det\bA=
a^2f'^2\det(\mat_{\alpha\beta})
+O(1/a) \ 
\end{equation}
and the inverse matrix 
 $\bAi$ when it is expressed as
function of $\mati$ and $\partial t$
takes the form 
\begin{eqnarray}\label{baif}
\bAi^{\alpha\beta}=\mati^{\alpha \beta} \ ,
\bAi^{x\beta}=\partial_\alpha t \mati^{\alpha \beta} \ ,
\nonumber \\
\bAi^{\alpha x}=\mati^{\alpha \beta}\partial_\beta t \ ,
\bAi^{xx}=\partial_\alpha t \mati^{\alpha \beta}
\partial_\beta t \ , \nonumber \\
\end{eqnarray}
where the relations in (\ref{baif}) hold up to
corrections of order $1/a^2$.

Now using the form of the matrix $\bA$ (\ref{bAf})
and the equation $\bAi^{\mu\nu}\bA_{\nu\rho}=
\delta^\mu_\rho$ we easily determine following
exact  relation 
\begin{equation}\label{bar}
\bAi^{\mu x}_S-\bAi^{\mu\alpha}_S
\partial_\alpha t=\frac{1}{a^2f'^2}
\left(\delta^{\mu}_x-\bAi^{xx}_Sg_{99}-\frac{1}{2}
\left(\bAi^{\mu\alpha}H_{\alpha x}+
H_{x\alpha}\bAi^{\alpha \mu}
\right)\right) \ . 
\end{equation}
Then with the help of (\ref{bar})
 we can write
the second term in (\ref{eqT}) as
\begin{eqnarray}
\partial_\mu \left[e^{-\Phi}V
\sqrt{-\det\bA}
 \bAi^{\mu\nu}_S\partial_\nu T
\right]=\nonumber \\
\partial_\mu \left[
e^{-\Phi}Vaf'\frac{1}{a^2f'^2}
(\delta^{\mu}_x -\bAi^{xx}_Sg_{99}-
\frac{1}{2}(\bAi^{\mu\alpha}H_{\alpha x}
+H_{x\alpha}\bAi^{\alpha \mu})
)\sqrt{-\det\mat}\right] \nonumber \\
\end{eqnarray}

Following \cite{Sen:2003tm} we can
now argue that due to the explicit
factor of $a^2f'^2$ in the 
denominator the 
leading contribution
from individual terms in this expression is now
of order $a$ and hence we 
can use the approximative
results of $\det\bA$ and $\bAi$
given in  (\ref{dea}) and (\ref{baif})
 to analyse the DBI part of the
 equation of motion
for tachyon (\ref{eqT})
\begin{eqnarray}\label{teqf}
\partial_\mu \left[
e^{-\Phi}V\sqrt{-\det\bA}af'\frac{1}{a^2f'^2}
(\delta^{\mu}_x -\bAi^{\mu x}g_{99}-
\frac{1}{2}\bAi^{\mu\alpha}H_{\alpha x}
-\frac{1}{2}H_{x\alpha}\bAi^{\alpha \mu}
)\right]-\nonumber \\
-e^{-\Phi}V'\sqrt{-\det\bA}=
\nonumber \\
\partial_x
\left[e^{-\Phi}V
\sqrt{-\det \mat}
(1-\mati^{\alpha\beta}_Sg_{99}\partial_\alpha t
\partial_\beta t
-\frac{1}{2}\partial_\beta t\mati^{\beta\alpha}H_{\alpha x}
-\frac{1}{2}H_{x\alpha}\mati^{\alpha\beta}\partial_\beta t 
)\right]-\nonumber \\
-\partial_\alpha
\left[e^{-\Phi}
V\sqrt{-\det\mat}\left(\mati^{\alpha\beta}_S
g_{99}\partial_\beta t
+\frac{1}{2}\mati^{\alpha\beta}H_{\beta x}
+\frac{1}{2}H_{x\alpha} \mati^{\alpha\beta}\right)
\right]-\nonumber \\
-af'e^{-\Phi}V'
\sqrt{-\det\mat}=\nonumber \\
=V\left\{\partial_x
\left[e^{-\Phi}\sqrt{-\det \mat}
(1-\mati^{\alpha\beta}_Sg_{99}\partial_\alpha t
\partial_\beta t-\right.\right.
\nonumber \\
\left.\left.-
\frac{1}{2}H_{x\alpha}
\mati^{\alpha\beta}\partial_\beta t
-\frac{1}{2}\partial_\alpha t
\mati^{\alpha\beta}H_{\beta x})
\right]-\right.\nonumber \\
\left. -\partial_\alpha
\left[e^{-\Phi}\sqrt{-\det \mat}
(\mati^{\alpha\beta}_Sg_{99}
\partial_\beta t+\frac{1}{2}
\mati^{\alpha\beta}
H_{\beta x}
+\frac{1}{2}H_{x\beta}\mati^{\beta\alpha}
)\right] \right\} \ .
\nonumber \\
\end{eqnarray}
We should now more carefully
interpret the result given
above. Firstly, as we know
from the previous section the
tachyon potential   $V$ is equal to zero for
$x-t(\xi)\neq 0$ while for $
x-t(\xi)=0$ we get $V(0)=\tau_p$ in
the limit $a\rightarrow \infty$.  Moreover,
we will show in the next subsection
that the tachyon current $J_T$
 is equal to $J_T=-V\tilde{J}_9$
when it is  evaluated
on the ansatz (\ref{ansTf})   and 
(\ref{ansAf}). Note that 
  $\tilde{J}_9$ 
is gauge fixed version of the
current (\ref{currentxkbps}). 
The main point is that the tachyon
equation of motion is obeyed for
$x-t(\xi)\neq 0$ while 
for $x=t(\xi)$ we should demand that
 the expression in the
bracket in (\ref{teqf}) together with
$-\tilde{J}_9$ should in be equal to zero.
If we now use the fact
that  
\begin{eqnarray}
H_{x\alpha}\mati^{\alpha\beta}\partial_\beta t
+\partial_\beta t \mati^{\beta \alpha}H_{\alpha x}=
2\mati^{\alpha\beta}_Sg_{I9}\partial_\alpha X^I
\partial_\beta t-2\mati^{\alpha\beta}_A
b_{I9}\partial_\alpha X^I\partial_\beta t \ , 
\nonumber \\
\mati^{\alpha\beta}
H_{\beta x}
+H_{x\beta}\mati^{\beta\alpha}=
2\mati^{\alpha\beta}_Sg_{I9}\partial_\beta X^I
-2\mati^{\alpha\beta}_A b_{9I}
\partial_\beta X^I \ 
\nonumber \\
\end{eqnarray}
we can write
the  expression in the bracket in
(\ref{teqf}) with $-\tilde{J}_9$ in the form 
\begin{eqnarray}\label{eqtf}
\frac{\delta e^{-\Phi}}
{\delta x}
\sqrt{-\det \mat}
+\frac{e^{-\Phi}}{2}
\left(\frac{\delta g_{MN}}
{\delta x}\partial_\alpha Y^M
\partial_\beta Y^N+
\frac{\delta b_{MN}}{\delta x}
\partial_\alpha Y^M\partial_\beta Y^N\right)
\mati^{\beta\alpha}
\sqrt{-\det\mat}-
\nonumber \\
-\partial_\alpha
\left[e^{-\Phi}
\sqrt{-\det \mat}
\mati^{\alpha\beta}_S
g_{9M}\partial_\beta Y^M\right]
-\partial_x\left[e^{-\Phi}
\sqrt{-\det\mat}\mati^{\alpha
\beta}_Sg_{9M}\right]\partial_\alpha t
\partial_\beta Y^M
-\nonumber \\
-\partial_\alpha\left[e^{-\Phi}
\sqrt{-\det \mat}\mati^{\alpha\beta}_A
b_{9M}\partial_\beta Y^M\right]
-\partial_9\left[e^{-\Phi}\sqrt{-\det\mat}
\mati^{\alpha\beta}_Ab_{9M}\right]
\partial_\alpha Y^M\partial_\beta t
-\tilde{J}_9=0 \ ,
\nonumber \\
\end{eqnarray}
where we have introduced the notation
\begin{equation}\label{notyt}
Y^M \ , M=0,\dots,9 \ ,
Y^I=X^I \ , I=0,\dots,8 \ , Y^9=t \ .
\end{equation}
It is important to stress that in (\ref{eqtf})
we  firstly perform
the derivative with respect to
 $x$ and
then we replace $x$ with  $t(\xi)$. 
  Then the presence  of the
following expressions in (\ref{teqf})
\begin{equation}
-\partial_x\left[e^{-\Phi}
\sqrt{-\det\mat}\mati^{\alpha
\beta}g_{9M}\right]\partial_\alpha t
\partial_\beta Y^M
-\partial_x\left[e^{-\Phi}\sqrt{-\det\mat}
\mati^{\alpha\beta}_Ab_{9M}\right]
\partial_\alpha Y^M\partial_\beta t
\end{equation}
 is crucial for  an
interpretation
of $t(\xi)$ as an additional 
scalar field that parametrises the
position of D(p-1)-brane in $x$ 
direction. 

To see this more clearly let us
compare  (\ref{eqtf})
  with the equation of motion  
(\ref{eqXbps}) for $K=9$ 
and observe that the
expression on the third
line in (\ref{eqXbps}) can
be written as
\begin{eqnarray}
\partial_\alpha\left[
e^{-\Phi}
\sqrt{-\det\mat}
g_{9M}\partial_\beta Y^M
\mati^{\beta\alpha}_{S}\right]=
\nonumber \\
=\partial_\alpha \left[
e^{-\Phi(\xi,x)}
\sqrt{-\det\mat (\xi,x)}
g_{9M}\partial_\beta Y^M
\mati^{\beta\alpha}_{S}(\xi,x)
\right]
\nonumber \\
+\partial_x
\left[e^{-\Phi(\xi,x)}
\sqrt{-\det\mat_{BPS}(\xi,x)}
g_{9M}\mati^{\beta\alpha}_{S}(\xi,x)
\right]
\partial_\alpha Y\partial_\beta Y^M \ ,
\nonumber \\
\end{eqnarray}
where on the second line the derivative
with respect to $\xi^\alpha$ 
treats $x$ as an independent variable
so that we firstly perform derivative with
respect to $\xi^\alpha$ and then 
we replace
$x$ with $Y$. We see that this
prescription coincides with the
expressions on the second line in
(\ref{eqtf}).  In the same
way we can proceed with the
expression on the fourth line
in (\ref{eqXbps}) 
\begin{eqnarray}
\nonumber \\
-\partial_\alpha
\left[e^{-\Phi}
\sqrt{-\det\mat}
b_{9M}\partial_\beta Y^M 
\mati^{\beta\alpha}_{A}
\right]=\nonumber \\
\partial_\alpha \left[
e^{-\Phi(\xi,x)}
\sqrt{-\det\mat(\xi,x)}
b_{9M}\partial_\beta Y^M
\mati^{\beta\alpha}_{A}(\xi,x)
\right]
\nonumber \\
+\partial_x
\left[e^{-\Phi(\xi,x)}
\sqrt{-\det\mat(\xi,x)}
b_{9M}\mati^{\beta\alpha}_{A}(\xi,x)
\right]
\partial_\alpha Y\partial_\beta Y^M \ \nonumber \\
\end{eqnarray}
and this again coincides with the
expressions on the fourth line in
(\ref{eqtf}).  In summary,
the location of the tachyon kink in
the $x^9$ direction 
is completely determined 
by field $t(\xi)$ that obeys the
equation of motion (\ref{eqXbps}) for
$K=9$. 
 
Now we come to the analysis of
the equation of motion for $X^K \ ,
K=0,\dots,8$.
For the ansatz (\ref{ansTf}) and
(\ref{ansAf}) the first term in (\ref{eqX})
takes the form
\begin{equation}\label{eqymp1}
\frac{\delta e^{-\Phi}}
{\delta X^K}
V\sqrt{-\det\bA}=
af'V\frac{\delta e^{-\Phi}}
{\delta X^K}\sqrt{-\det\mat} \ .
\end{equation}
On the other hand the expression
on the  second line 
in (\ref{eqX}) can be written as
\begin{eqnarray}\label{eqymp2}
e^{-\Phi}V\left[
\frac{\delta g_{MN}}
{\delta X^K}\partial_\mu X^M
\partial_\nu X^N
+\frac{\delta b_{MN}}
{\delta X^K}
\partial_\mu X^M\partial_\nu X^N\right]
\bAi^{\nu\mu}
\sqrt{-\det\bA}=\nonumber \\
af'Ve^{-\Phi}
\left[\frac{\delta g_{MN}}
{\delta Y^K}\partial_\alpha Y^M
\partial_\beta Y^N
+\frac{\delta b_{MN}}
{\delta Y^K}
\partial_\alpha Y^M\partial_\beta Y^N\right]
\mati^{\beta\alpha}\sqrt{-\det\mat} \  ,  
\nonumber \\
\end{eqnarray}
where we have used the notation
(\ref{notyt}). 
Finally we will analyse the
 expression 
on the third and the fourth line in 
(\ref{eqX}) that can be written
as
\begin{equation}\label{pbg}
\partial_\mu\left[
e^{-\Phi}V(g_{KM}
\partial_\nu X^M
\bAi^{\nu\mu}_S+
b_{KM}
\partial_\nu X^M 
\bAi^{\nu\mu}_A)
\sqrt{-\det\bA}\right] \ . 
\end{equation}
After some length calculations
 we
obtain that (\ref{pbg})
for the ansatz (\ref{ansTf})
and (\ref{ansAf}) takes
the form
\begin{eqnarray}\label{pxf}
aVf'\left(
\partial_x\left[e^{-\Phi}(
g_{KM}\mati^{\alpha\beta}_S+
b_{KM}\mati^{\alpha\beta}_A)
\sqrt{-\det\mat}\right]
\partial_\alpha 
Y^M\partial_\beta t
\right.
\nonumber \\
\left.+
\partial_\alpha\left[
e^{-\Phi}(g_{KM}\partial_\beta 
Y^M\mati^{\beta\alpha}_S+
b_{KM}\partial_\beta Y^M
\mati^{\beta\alpha}_A)
\sqrt{-\det\mat}\right]
\right) \ . 
\nonumber \\
\end{eqnarray}
Finally, using 
(\ref{eqymp1}), (\ref{eqymp2}) and
(\ref{pxf}) we get
\begin{eqnarray}\label{ebr}
af'V\left\{-\frac{\delta e^{-\Phi}}
{\delta X^K}\sqrt{-\det\mat} 
+\right. \nonumber \\
\left.-\frac{e^{-\Phi}}{2}
\left[\frac{\delta g_{MN}}
{\delta Y^K}\partial_\alpha Y^M\partial_\beta Y^N
+\frac{b_{MN}}
{\delta Y^K}
\partial_\alpha Y^M\partial_\beta Y^N\right]
\mati^{\beta\alpha}\sqrt{-\det\mat}\right.  
\nonumber \\
\left.+\partial_x\left[e^{-\Phi}(
g_{KM}\mati^{\alpha\beta}_S+
b_{KM}\mati^{\alpha\beta}_A)
\sqrt{-\det\mat}\right]
\partial_\alpha Y^M
\partial_\beta t
\right.
\nonumber \\
\left.+
\partial_\alpha\left[
e^{-\Phi}(g_{KM}\partial_\beta Y^M
\mati^{\beta\alpha}_S+
b_{KM}\partial_\beta Y^M
\mati^{\beta\alpha}_A)
\sqrt{-\det\mat}\right]
+\tilde{J}_K\right\}=0 \ 
\nonumber \\
\end{eqnarray}
using  the result that
will be proven in the 
next subsection that
the current $J_K$ is 
equal to $aVf'\tilde{J}_K$, where
 $\tilde{J}_K$ is given in (\ref{currentxkbps}).

As we know from the previous
section the expression $af'V$
goes to zero in
the limit $a\rightarrow \infty$ when
 $x\neq t(\xi)$. On the other hand
for  $x=t(\xi)$ the potential
$V(0)=\tau_p$ for arbitrary $a$ 
 and hence   in order to obey the
equation of motion for $X^K$ (\ref{eqX})
 we get that the expression
in the bracket $\left\{\dots\right\}$
should vanish for $x=t(\xi)$.
However this is precisely the
equation of motion (\ref{eqXbps})
and hence we again obtain the
result that the scalar modes $X^K$
should solve the equation of motion
that arise from the action for
BPS D(p-1)-brane.

Since we mean that it is very
important to find the correct 
interpretation of the equation
(\ref{ebr}) we we would like
again   stress that 
in the  expression in the bracket in
(\ref{ebr}) 
 we firstly  perform   a derivative
with respect to 
$\xi^{\alpha}$ and then 
 we  replace $x$ with $t(\xi)$ 
in the limit
$a\rightarrow \infty$. 
This fact  implies that 
$t(\xi)$ is  an scalar mode
that parametrises the location
of D(p-1)-brane in the
$x^9$ direction.

To complete the discussion of the
equation of motion for $X^K$ we should
also analyse the equation of motion for
$X^9$. If we proceed in the same
way as for $X^K$ that was analysed
above we obtain that the equation
of motion for $X^9$ takes the form 
\begin{eqnarray}\label{eqx9}
af'V\left(
-\frac{e^{-\Phi}}{2}
\sqrt{-\det\mat}\left[\partial_x g_{MN}
+\partial_x b_{MN}\right]\partial_\alpha Y^M
\partial_\beta Y^N\mati^{\beta\alpha} 
-\partial_x[e^{-\Phi}]\sqrt{-\det \mat}+
\right. \nonumber \\ 
\left.+\partial_x\left[e^{-\Phi}g_{9M}\mati^{\alpha\beta}_S
\sqrt{-\det\mat}\right]\partial_\alpha t\partial_\beta X^M+
\partial_\beta\left[e^{-\Phi}g_{9M}\partial_\alpha Y^M
\mati^{\alpha\beta}_S
\sqrt{-\det\mat}\right]+\right. \nonumber \\
\left.\partial_x\left[e^{-\Phi}b_{9M}\mati^{\beta\alpha}_A
\sqrt{-\det\mat}\right]\partial_\alpha t\partial_\beta X^M
+\partial_\beta
\left[e^{-\Phi}b_{9M}\partial_\alpha X^M
\mati^{\alpha\beta}_A\sqrt{-\det\mat}\right]
+\tilde{J}_9\right)=0 \ , 
\nonumber \\
\end{eqnarray}
where we have again used the result from 
the next subsection that 
 $J_9=afV'\tilde{J}_9$. 
We see that the expression in the
bracket $\left(\dots\right)$ in 
(\ref{eqx9}) coincides with the equation
of motion (\ref{eqXbps}) for $K=9$. 
This is nice result since we should obtain
ten independent 
equations for scalar modes
and we see that the equation of motion
for $T$ and for $X^9$ imply  one
equation of motion for  mode $t$. 

Finally we come to the analysis
of the  equation of motion for
$A_\mu$ given in
(\ref{eqA}). For $\mu=\alpha$
 the
 DBI part of the equation
of motion (\ref{eqA}) 
takes the form
\begin{eqnarray}\label{eqadbi}
\partial_x[Vaf']e^{-\Phi}
\mati^{\alpha\beta}_A
\partial_\beta t\sqrt{-\det\mat}
+\partial_\beta[Vaf']e^{-\Phi}
\mati^{\alpha\beta}_A
\sqrt{-\det\mat}+\nonumber \\
Vaf'\left(\partial_x
\left[e^{-\Phi}
\mati^{\alpha\beta}_A
\sqrt{-\det\mat}\right]\partial_\beta t+
\partial_\beta\left[e^{-\Phi}\mati^{\alpha
\beta}_A\sqrt{-\det\mat}\right]\right) \ . 
\nonumber \\
\end{eqnarray}
Again using (\ref{pdvf}) we see
that the expressions on the first
line cancel.  If we now combine
(\ref{eqadbi}) with   
(\ref{gcf}) we obtain final form
of the equation of motion for
$A_\alpha$
\begin{eqnarray}\label{eqAaf}
Vaf'\left(\partial_x
\left[e^{-\Phi}
\mati^{\alpha\beta}_A
\sqrt{-\det\mat}\right]\partial_\beta t+
\partial_\beta\left[e^{-\Phi}\mati^{\alpha
\beta}_A\sqrt{-\det\mat}\right]+
\tilde{J}^{\alpha}
\right)=0 \ . 
\nonumber \\
\end{eqnarray}
As usual we demand that
 the expression
in the bracket $\left(\dots\right)$
in (\ref{eqAaf})  should be equal to zero for
$x=t(\xi)$.  Then the vanishing
of this expression is equivalent to
\begin{equation}\label{eqfa}
\partial_\alpha\left[
e^{-\Phi(t(\xi))}\sqrt{-\det\mat
(t(\xi))}\mati^{\beta\alpha}_A(t(\xi))
\right]+\tilde{J}^\alpha(t(\xi))=0 
\end{equation}
that is an  equation
of motion for the gauge field  
given in (\ref{eqAbps}). 

Finally, the DBI part of the equation of
motion (\ref{eqA}) for $\mu=x$
and for the ansatz (\ref{ansTf}),
(\ref{ansAf})
takes the form
\begin{eqnarray}\label{pux}
\partial_\nu \left[V
e^{-\Phi}\bAi_A^{x\nu}
\sqrt{-\det\bA}\right]=
\partial_\beta\left[Vaf'e^{-\Phi}
\partial_\alpha 
t \mati^{\alpha\beta}_A
\sqrt{-\det\mat}\right]=
\nonumber \\
=Vaf'\partial_\beta t
\partial_\alpha\left[e^{-\Phi}
\mati^{\beta\alpha}_A\sqrt{-\det\mat}\right] 
\  \nonumber \\
\end{eqnarray}
using (\ref{pdvf}) and then an
antisymmetry of 
the matrix
$\mati^{\alpha\beta}_A$. 
Now with the help of the current $J^x$ given
in (\ref{jxpf}) and with (\ref{pux})
the equation of motion 
(\ref{eqA}) for $\mu=x$  takes the form
\begin{eqnarray}
aVf'\left\{
\partial_\beta t\left(
\partial_\alpha\left[e^{-\Phi}
\mati^{\beta\alpha}_A\sqrt{-\det\mat}\right]
+\tilde{J}^{\beta}\right)\right\}=
  \nonumber \\
=aVf'\left\{
\partial_\beta t\left(
\partial_\alpha\left[e^{-\Phi}
\mati^{\beta\alpha}_A\sqrt{-\det\mat}\right]
+\partial_x
\left[e^{-\Phi}
\mati^{\alpha\beta}_A
\sqrt{-\det\mat}\right]\partial_\alpha t
+\tilde{J}^{\beta}\right)\right\}=0
\ ,  \nonumber \\
\end{eqnarray}
where we have included an expression
$aVf'\partial_\beta t\partial_{\alpha}t\partial_x
\left[e^{-\Phi}
\mati^{\alpha\beta}_A
\sqrt{-\det\mat}\right]$ that
vanishes thanks to the antisymmetry
of $\mati_A^{\alpha\beta}$ however
whose presence is crucial for an interpretation
of $t$ as an embedding coordinate. 
Following arguments given above we 
obtain that the expression in the bracket
$\left\{\dots\right\}$ should be equal to
zero for $x=t(\xi)$ in the limit $a\rightarrow
\infty$. We see that this 
holds since as we have
argued above the massless modes 
obey (\ref{eqAbps}). 

In summary, we have shown that the
dynamics of the tachyon kink is governed
by the equation of motion that arises from the
DBI and WZ action for D(p-1)-brane that
is localised at the point $x=t(\xi)$.  To really
conclude this section we should now
evaluate currents $J_M, J^\mu$ and 
$J^T$. 
\subsection{Analysis of currents}
In this subsection we will
analyse the currents 
(\ref{currentA}) , (\ref{currentT}) and
(\ref{currentXK}) for the ansatz
given in (\ref{ansTf}) and (\ref{ansAf}). 
We will see that 
this analysis is much more difficult
that in the case when we did not
impose any gauge fixing conditions. 

We start with the gauge current
(\ref{currentA}) where  $\mu_1=\alpha_1$. 
In this case we
 get
\begin{eqnarray}\label{ja1}
J^{\alpha_1}=
\sum_{n\geq 0}
\frac{2n}{n! 2^nq!}
\epsilon^{\alpha_1\mu_1\dots \mu_{p+1}}
\partial_{\mu_2}\left[V(T) (\mF)^{n-1}_{\mu_3\dots \mu_{2n}}
C_{\mu_{2n+1}\dots \mu_{p}}\partial_{\mu_{p+1}}T
\right]=\nonumber \\
=\sum_{n\geq 0}
\frac{2n}{n! 2^nq!}
\epsilon^{\alpha_1\alpha_2\dots \alpha_p x}
\partial_{\alpha_2}
\left[V(T) (\mF)^{n-1}_{\alpha_3\dots \alpha_{2n}}
C_{\alpha_{2n+1}\dots \alpha_{p}}\partial_{x}T
\right]+\nonumber \\
\sum_{n\geq 0}
\frac{2n}{n! 2^nq!}
\epsilon^{\alpha_1x\alpha_2\dots \alpha_p }
\partial_{x}\left[V(T) (\mF)^{n-1}_{\alpha_2\dots \alpha_{2n}}
C_{\alpha_{2n+1}\dots \alpha_{p-1}}\partial_{\alpha_p}T
\right]+\nonumber \\
+\sum_{n\geq 0}
\frac{4n(n-1)}{n! 2^nq!}
\epsilon^{\alpha_1\alpha_2x\alpha_4\dots \alpha_p\alpha_3 }
\partial_{\alpha_2}
\left[V(T)\mF_{x\alpha_3} (\mF)^{n-2}_{
\alpha_5\dots \alpha_{2n}}
C_{\alpha_{2n+1}\dots \alpha_{p}}\partial_{\alpha_3}T
\right]+ \nonumber \\
+\sum_{n\geq 0}\frac{2nq}{n! 2^nq!}
\epsilon^{\alpha_1\dots\alpha_{2n}x
\alpha_{2n+2}\dots\alpha_p\alpha_{2n+1}}
\partial_{\alpha_2}
\left[V(T)(\mF)^{n-1}_
{\alpha_3\dots\alpha_{2n}}
C_{x\alpha_{2n+2}\dots \alpha_{p}}
\partial_{\alpha_{2n+1}}T
\right] \ .  \nonumber \\
\end{eqnarray}
It can be shown that for the
ansatz (\ref{ansTf})
 the expressions on the second 
  and the  third line 
in (\ref{ja1}) take the form 
\begin{eqnarray}\label{pc1}
af'V\sum_{n\geq 0}
\frac{2n}{n! 2^nq!}
\epsilon^{\alpha_1\dots \alpha_p x}
\partial_{\alpha_2}
\left[ (\mF)^{n-1}_{\alpha_3\dots \alpha_{2n}}
C_{\alpha_{2n+1}\dots \alpha_{p}}
\right]+
\nonumber \\ 
+aVf'\sum_{n\geq 0}
\frac{2n}{n! 2^nq!}
\epsilon^{\alpha_1\dots\alpha_px}
\partial_{x}
\left[ (\mF)^{n-1}_{\alpha_3\dots \alpha_{2n}}
C_{\alpha_{2n+1}\dots \alpha_{p}}
\right]\partial_{\alpha_2}t \ . 
\nonumber \\
\end{eqnarray}
Now we come to one important point.
As we know from the previous section the
factor $aVf'$
vanishes for $x\neq t(\xi)$
for $a\rightarrow \infty$. At
the same time we argued
that  we should regard $t(\xi)$ as
an embedding coordinate. 
On the other
hand $\mF_{\alpha\beta}$ contains 
an embedding of
$B$ that is equal to
\begin{equation}
B_{IJ}\partial_{\alpha}X^I\partial_\beta X^J
\end{equation}
and we also have
\begin{equation}
C_{\alpha_{2n+1}\dots \alpha_p}=
C_{I_{2n+1}\dots I_p}
\partial_{\alpha_{2n+1}}X^{I_{2n+1}}
\dots \partial_{\alpha_p}X^{I_p} \ .
\end{equation}
Now we would like to argue that whenever 
some term in any current
will contain a  factor $\partial_{\alpha_x}
t$ we can replace all $\mF_{\alpha\beta}$
 and all $C_{\alpha_{2n+1}\dots \alpha_p}$ with
$\tilde{\mF}_{\alpha\beta}$ and 
$\tilde{C}_{\alpha_{2n+1}\dots \alpha_p}$ 
where
\begin{eqnarray}\label{dfct}
\tilde{\mF}_{\alpha\beta}=\mF_{\alpha\beta}+
B_{MN}\partial_{\alpha}Y^M\partial_{\beta}Y^N
\ , \nonumber \\
\tilde{C}_{\alpha_{2n+1}\dots \alpha_p}=
C_{M_{2n+1}\dots M_p}
\partial_{\alpha_{2n+1}}Y^{M_{2n+1}}
\dots \partial_{\alpha_p}Y^{M_p} \ ,  
\nonumber \\ 
\end{eqnarray}
where $Y^M$ was introduced in (\ref{notyt}).
To see that this replacement is
correct note that  the
 additional terms  in  expressions (We 
mean  expressions with the
overall multiplicative
factor  $\partial_{\alpha_x}t$)  , when
we  replace $\mF$ with $\tilde{\mF}$
and $C$ with $\tilde{C}$ 
contain derivative of $t$ in the form $\partial_{\alpha_y}t$. 
Now thanks to the existence of the factor
$\epsilon^{\alpha_1\dots\alpha_p x}$ it is clear that
these terms after multiplication with $\partial_{\alpha_x}t$
vanish since
\begin{equation}
\epsilon^{\alpha_1 \dots \alpha_x \dots
\alpha_y \dots \alpha_p x}
\partial_{\alpha_x}t\partial_{\alpha_y}t=
0 \ . 
\end{equation} 
Now we proceed to the analysis 
of the expression
on the fourth   line in (\ref{ja1})
\begin{eqnarray}\label{pc2}
\sum_{n\geq 0}
\frac{4n(n-1)}{n! 2^nq!}
\epsilon^{\alpha_1\alpha_2x\alpha_4\dots \alpha_p\alpha_3 }
\partial_{\alpha_2}
\left[V(T)\mF_{x\alpha_4} (\mF)^{n-2}_{
\alpha_5\dots \alpha_{2n}}
C_{\alpha_{2n+1}\dots \alpha_{p}}\partial_{\alpha_3}T
\right]= \nonumber \\
aVf'\sum_{n\geq 0}
\frac{4n(n-1)}{n! 2^nq!}
\epsilon^{\alpha_1\dots\alpha_p x}
\partial_{\alpha_2}
\left[b_{9I}\partial_{\alpha_3}t
\partial_{\alpha_4}X^I (\mF)^{n-2}_{
\alpha_5\dots \alpha_{2n}}
C_{\alpha_{2n+1}\dots \alpha_{p}}
\right] \nonumber \\
\end{eqnarray}
using the fact that
\begin{equation}
\mF_{x\alpha}=F_{x\alpha}+b_{9I}\partial_{\alpha}X^I=
b_{9I}\partial_{\alpha}X^I \ .
\end{equation}
Now it is easy to see that 
(\ref{pc2}) together  with (\ref{pc1}) 
gives
\begin{eqnarray}\label{pc12f}
af'V\sum_{n\geq 0}
\frac{2n}{n! 2^nq!}
\epsilon^{\alpha_1\dots \alpha_p x}
\partial_{\alpha_2}
\left[ (\mF)^{n-1}_{\alpha_3\dots \alpha_{2n}}
C_{\alpha_{2n+1}\dots \alpha_{p}}
\right]+
\nonumber \\ 
+aVf'\sum_{n\geq 0}
\frac{2n}{n! 2^nq!}
\epsilon^{\alpha_1\dots\alpha_px}
\partial_{x}
\left[ (\mF)^{n-1}_{\alpha_3\dots \alpha_{2n}}
C_{\alpha_{2n+1}\dots \alpha_{p}}
\right]\partial_{\alpha_2}t + 
\nonumber \\
+aVf'\sum_{n\geq 0}
\frac{4n(n-1)}{n! 2^nq!}
\epsilon^{\alpha_1\dots\alpha_p x}
\partial_{\alpha_2}
\left[b_{9I}\partial_{\alpha_3}t
\partial_{\alpha_4}X^I (\mF)^{n-2}_{
\alpha_5\dots \alpha_{2n}}
C_{\alpha_{2n+1}\dots \alpha_{p}}
\right]= \nonumber \\
=af'V\sum_{n\geq 0}
\frac{2n}{n! 2^nq!}
\epsilon^{\alpha_1\dots \alpha_p x}
\partial_{\alpha_2}
\left[ (\tilde{\mF})^{n-1}_{\alpha_3\dots \alpha_{2n}}
C_{\alpha_{2n+1}\dots \alpha_{p}}
\right]+
\nonumber \\ 
+aVf'\sum_{n\geq 0}
\frac{2n}{n! 2^nq!}
\epsilon^{\alpha_1\dots\alpha_px}
\partial_{x}
\left[ (\tilde{\mF})^{n-1}_{\alpha_3\dots \alpha_{2n}}
\tilde{C}_{\alpha_{2n+1}\dots \alpha_{p}}
\right]\partial_{\alpha_2}t \ .
\nonumber \\
\end{eqnarray}
To complete the discussion of the
current we should analyse the expression 
on the last line in (\ref{ja1})
\begin{eqnarray}\label{pc3}
\sum_{n\geq 0}\frac{2nq}{n! 2^nq!}
\epsilon^{\alpha_1\dots\alpha_{2n}x
\alpha_{2n+2}\dots\alpha_p\alpha_{2n+1}}
\partial_{\alpha_2}
\left[V(T)(\mF)^{n-1}_
{\alpha_3\dots\alpha_{2n}}
C_{x\alpha_{2n+2}\dots \alpha_{p}}
\partial_{\alpha_{2n+1}}T
\right]=\nonumber \\
aVf'\sum_{n\geq 0}\frac{2nq}{n! 2^nq!}
\epsilon^{\alpha_1\dots\alpha_p x}
\partial_{\alpha_2}
\left[(\mF)^{n-1}_
{\alpha_3\dots\alpha_{2n}}
C_{x\alpha_{2n+2}\dots \alpha_{p}}
\partial_{\alpha_{2n+1}}t
\right] \ . \nonumber \\ 
\end{eqnarray}
Now we see that  (\ref{pc3}) is precisely
the expression that is needed to replace
 $C_{\alpha_{2n+1}\dots\alpha_p}$
with $\tilde{C}_{\alpha_{2n+1}\dots\alpha_p}$ in
(\ref{ja1}). Finally, if we combine (\ref{pc12f})
with (\ref{pc3}) we obtain following form
of the current $J^{\alpha_1}$
\begin{eqnarray}\label{gcf}
J^{\alpha_1}=af'V
\sum_{n\geq 0}
\frac{2n}{n! 2^nq!}
\epsilon^{\alpha_1\alpha_2\dots \mu_{p}x}
\partial_{\alpha_2}
\left[(\tilde{\mF})^{n-1}_{\alpha_3\dots \alpha_{2n}}
\tilde{C}_{\alpha_{2n+1}
\dots \alpha_{p}}\right]
+\nonumber \\
+aVf'\sum_{n\geq 0}
\frac{2n}{n! 2^nq!}
\epsilon^{\alpha_1\dots\alpha_px}
\partial_{x}
\left[ (\tilde{\mF})^{n-1}_{\alpha_3\dots \alpha_{2n}}
\tilde{C}_{\alpha_{2n+1}\dots \alpha_{p}}
\right]\partial_{\alpha_2}t 
=
afV\tilde{J}^{\alpha_1} \ ,  \nonumber \\
\end{eqnarray}
where $\tilde{J}^{\alpha_1}$ is a gauge field
current for D(p-1)-brane given 
in (\ref{currentAbps}). Note also that
the term on the second line in (\ref{gcf})
is exactly the right one in order to
interpret $t$ as an embedding coordinate
since in the expression on the first line
in (\ref{gcf}) the 
partial derivative $\partial_{\alpha_2}$
treats $x$ as an independent variable. 
 We will also see that in all
other currents  similar  additional terms
appear as well.

Finally, we will analyse the gauge
current for $\mu_1=x$
\begin{eqnarray}\label{jxp}
J^{x}=
\sum_{n\geq 0}
\frac{2n}{n! 2^nq!}
\epsilon^{x\alpha_2\alpha_3\dots \alpha_p\alpha_1}
\partial_{\alpha_2}
\left[V (\mF)^{n-1}_{\alpha_3\dots \alpha_{2n}}
C_{\alpha_{2n+1}\dots \alpha_{p}}\partial_{\alpha_1}T
\right]=\nonumber \\
=af'V\partial_{\alpha_1}t\sum_{n\geq 0}
\frac{2n}{n! 2^nq!}
\epsilon^{\alpha_1\dots\alpha_px}
\partial_{\alpha_2}
\left[(\mF)^{n-1}_{\alpha_3\dots \alpha_{2n}}
C_{\alpha_{2n+1}\dots \alpha_{p}}\right]
 \  , \nonumber \\
\end{eqnarray}
where we have used an antisymmetry
of $\epsilon^{x\alpha_1\dots \alpha_p}$ under exchange
of $\alpha_1$ and $\alpha_p$ so that
$\epsilon^{\alpha_1\dots\alpha_p}\partial_{\alpha_1}
\partial_{\alpha_p}t=0$. 

Thanks to the presence of the term 
$\partial_{\alpha_1}t$ we
can, following discussion 
given above, everywhere replace
$\mF$  with $\tilde{\mF}$ 
 and $C$ with $\tilde{C}$. From the
same reason we can  add to (\ref{jxp}) 
an expression $aVf'\sum_{n\geq 0}
\frac{2n}{n! 2^nq!}
\epsilon^{\alpha_1\dots\alpha_px}
\partial_{x}
\left[ (\tilde{\mF})^{n-1}_{\alpha_3\dots \alpha_{2n}}
\tilde{C}_{\alpha_{2n+1}\dots \alpha_{p}}
\right]\partial_{\alpha_2}t\partial_{\alpha_1}t$
that formally vanishes however with this term the
current (\ref{jxp}) can be written as
\begin{eqnarray}\label{jxpf}
J^x=af'V\partial_{\alpha_1}t\sum_{n\geq 0}
\frac{2n}{n! 2^nq!}
\epsilon^{\alpha_1\dots\alpha_px}
\partial_{\alpha_2}
\left[(\tilde{\mF})^{n-1}_{\alpha_3\dots \alpha_{2n}}
\tilde{C}_{\alpha_{2n+1}\dots \alpha_{p}}\right]+
 \nonumber \\
+aVf'\sum_{n\geq 0}
\frac{2n}{n! 2^nq!}
\epsilon^{\alpha_1\dots\alpha_px}
\partial_{x}
\left[ (\tilde{\mF})^{n-1}_{\alpha_3\dots \alpha_{2n}}
\tilde{C}_{\alpha_{2n+1}\dots \alpha_{p}}
\right]\partial_{\alpha_2}t\partial_{\alpha_1}t
=aVf'\partial_{\alpha_1}t\tilde{J}^{\alpha_1} \ . 
\nonumber \\
\end{eqnarray}
Let us now proceed to the analysis of
the tachyon current (\ref{currentT})
for the ansatz (\ref{ansTf}) and
(\ref{ansAf})
\begin{eqnarray}\label{jtr}
J_T=-V(T)\sum_{n\leq 0}
\frac{1}{n!(2!)^nq!}
\epsilon^{\mu_1\dots \mu_{p+1}}
\partial_{\mu_{p+1}}
\left[(\mF)^n_{\mu_1\dots \mu_{2n}}
C_{\mu_{2n+1}\dots \mu_p}\right]
\ . 
\nonumber \\
\end{eqnarray}
Now we  split  the calculations 
into  two 
parts, the first one when $\mu_{p+1}=x$ and
the second one when $\mu_{p+1}\neq x$. In
the first case we get
\begin{eqnarray}\label{pt1}
-V(T)\sum_{n\leq 0}
\frac{1}{n!(2!)^nq!}
\epsilon^{\alpha_1\dots \alpha_{p}
x}
\partial_{x}
\left[(\mF)^n_{\alpha_1\dots \alpha_{2n}}
C_{\alpha_{2n+1}
\dots \alpha_p}\right]=
\nonumber \\
=-V\sum_{n\leq 0}\frac{n}{n!(2!)^nq!}
\epsilon^{\alpha_1\dots \alpha_{p}
x}
\partial_{x}b_{IJ}\partial_{\alpha_1}X^I
\partial_{\alpha_2}X^J
\left[(\mF)^{n-1}_{\alpha_3\dots \alpha_{2n}}
C_{\alpha_{2n+1}
\dots \alpha_p}\right]-
\nonumber \\
-V\sum_{n\leq 0}\frac{q}{n!(2!)^nq!}
\epsilon^{\alpha_1\dots \alpha_{p}
x}
\left[(\mF)^{n}_{\alpha_1\dots \alpha_{2n}}
\partial_xC_{I_{2n+1}\dots I_p}
\partial_{\alpha_{2n+1}}
X^{I_{2n+1}}
\dots \partial_{\alpha_p}X^{I_p}
\right] \ . 
\nonumber \\
\end{eqnarray}
Now we  extend  the expression 
$\partial_xb_{IJ}
\partial_{\alpha_1}X^I
\partial_{\alpha_2}X^J$ 
as
\begin{eqnarray}
\partial_xb_{IJ}\partial_{\alpha_1}X^I
\partial_{\alpha_2}X^J+
\partial_xb_{9J}
\partial_{\alpha_1}t\partial_{\alpha_2}X^J+
\partial_xb_{I9}\partial_{\alpha_1}X^I
\partial_{\alpha_2}t-\nonumber \\
(\partial_xb_{9J}\partial_{\alpha_1}t
\partial_{\alpha_2}X^J+
\partial_xb_{I9}\partial_{\alpha_1}
X^I\partial_{\alpha_2}t) \ .
\nonumber \\
\end{eqnarray}
In the same way we can
proceed with the expression 
 $\partial_x C_{I_{2n+1}\dots I_p}$. Then the
expression (\ref{pt1}) 
 takes the form
\begin{eqnarray}\label{pt1f}
-V\sum_{n\leq 0}\frac{1}{n!(2!)^nq!}
\epsilon^{\alpha_1\dots \alpha_{p}x}
\partial_{x}b_{MN}
\partial_{\alpha_1}Y^M
\partial_{\alpha_2}Y^N
\left[(\mF)^{n-1}_{\alpha_3\dots \alpha_{2n}}
C_{\alpha_{2n+1}\dots \alpha_p}\right]-
\nonumber \\
-V\sum_{n\leq 0}\frac{q}{n!(2!)^nq!}
\epsilon^{\alpha_1\dots \alpha_{p}
x}
\left[(\mF)^{n}_{\alpha_1\dots \alpha_{2n}}
\partial_xC_{M_{2n+1}\dots M_p}
\partial_{\alpha_{2n+1}}Y^{M_{2n+1}}
\dots \partial_{\alpha_p}Y^{M_p}\right]+
\nonumber \\
+V\sum_{n\leq 0}\frac{2n}{n!(2!)^nq!}
\epsilon^{\alpha_1\dots \alpha_{p}x}
[\partial_xb_{9N}\partial_{\alpha_1}t
\partial_{\alpha_2}Y^N](\tilde{\mF})^{n-1}_{\alpha_1\dots 
\alpha_{2n}}
\tilde{C}_{\alpha_{2n+1}\dots \alpha_p}
\nonumber \\
+V\sum_{n\leq 0}
\frac{q}{n!(2!)^nq!}
\epsilon^{\alpha_1\dots\alpha_px}
(\mF)^n_{\alpha_1\dots\alpha_{2n}}
[\partial_xC_{9I_{2n+2}\dots I_p}
\partial_{\alpha_{2n+1}}t
\partial_{\alpha_{2n+2}}Y^I\dots
\partial_{\alpha_p}
Y^{I_p}] \ , 
\nonumber \\
\end{eqnarray}
where we have included 
tilde components defined in (\ref{dfct}).
We have also used the fact
that  we can write
$b_{9I}\partial_{\alpha_2}X^I=
b_{9M}\partial_{\alpha}Y^M$ and in the
same  way we can 
extend the embedding $C_{9I_{2n+2}
\dots I_p}\partial_{\alpha_{2n+2}}X^{I_{2n+2}}\dots
\partial_{\alpha_p}X^{I_{p}}$
to $C_{9M_{2n+2}
\dots M_p}\partial_{\alpha_{2n+2}}Y^{M_{2n+2}}\dots
\partial_{\alpha_p}Y^{M_{p}}$
using antisymmetry of $C_{M_1\dots M_q}$.

Let us now  consider the
case when  $\mu_{p+1}\neq x$ in
  (\ref{jtr}). In this
case we get
\begin{eqnarray}\label{pt2}
-V(T)\sum_{n\leq 0}
\frac{2n}{n!(2!)^nq!}
\epsilon^{x\alpha_2\dots \alpha_{p}\alpha_1}
\partial_{\alpha_1}\left[\mF_{x\alpha_2}
(\mF)^{n-1}_{\alpha_3\dots \alpha_{2n}}
C_{\alpha_{2n+1}\dots \alpha_p}\right]-
\nonumber \\
-V(T)\sum_{n\leq 0}
\frac{q}{n!(2!)^nq!}
\epsilon^{\alpha_1\dots \alpha_{2n}
x\alpha_{2n+2}\alpha_p\alpha_{2n+1}}
\partial_{\alpha_{2n+1}}
\left[(\mF)^{n}_{\alpha_1\dots \alpha_{2n}}
C_{x\alpha_{2n+2}\dots \alpha_p}\right]=
\nonumber \\
=V(T)\sum_{n\leq 0}
\frac{2n}{n!(2!)^nq!}
\epsilon^{\alpha_1\alpha_2\dots \alpha_{p}x}
\partial_{\alpha_1}\left[b_{9I}\partial_{\alpha_2}X^I
(\mF)^{n-1}_{\alpha_3\dots \alpha_{2n}}
C_{\alpha_{2n+1}\dots \alpha_p}\right]+
\nonumber \\
+V(T)\sum_{n\leq 0}
\frac{q}{n!(2!)^nq!}
\epsilon^{\alpha_1\dots
 \alpha_{p}x}
\partial_{\alpha_{2n+1}}
\left[(\mF)^{n}_{\alpha_1\dots \alpha_{2n}}
C_{9I_{2n+2}\dots I_p}
\partial_{\alpha_{2n+2}}X^{I_{2n+2}}\dots
\partial_{\alpha_p}X^{I_{p}}
\right]
\nonumber \\
\end{eqnarray}
using the fact that 
$F_{x\alpha_1}=F_{\alpha_1x}=0$. 

We will again  
argue that   terms 
 written on the  
third and the fourth line in
(\ref{pt1f})  
are important for an 
interpretation of $t$ as
an  embedding coordinate. 
In fact, following discussion
performed in previous section
it is easy to see that
\begin{eqnarray}\label{pex}
\partial_{\alpha_1}[b_{9I}\partial_{\alpha_2}X^I]=
\left.\partial_{\alpha_1}(b_{9I}(x,X))
\partial_{\alpha_2}X^I
\right|_{x=t(\xi)}+
\left.\partial_{x}(b_{9I}(x,X))
\partial_{\alpha_2}X^I
\right|_{x=t(\xi)}\partial_{\alpha_1}t
\nonumber \\
+b_{9I}\partial_{\alpha_1}\partial_{\alpha_2}X^I
\nonumber \\ \ , 
\end{eqnarray}
where the second term vanishes after multiplying this
derivative with $\epsilon^{\alpha_1\alpha_2\dots}$. 
In the same way we can show 
that the derivative 
$\partial_{\alpha_2}\mF_{\alpha_3\alpha_4}$
takes the form
\begin{eqnarray}
\partial_{\alpha_2}F_{\alpha_3\alpha_4}+
\left.\partial_x b_{IJ}\partial_{\alpha_3}X^I
\partial_{\alpha_4}X^J
\partial_{\alpha_2}t\right|_{x=t(\xi)}+\nonumber \\
+\left.\partial_{\alpha_2} b_{IJ}(x,X)
\partial_{\alpha_3}X^I
\partial_{\alpha_4}X^J\right|_{x=t(\xi)}+
b_{IJ}(x,X)\partial_{\alpha_2}
\left(\partial_{\alpha_3}X^I
\partial_{\alpha_4}X^J\right) \nonumber \\
\end{eqnarray}
If we multiply the expression given above with
$\epsilon^{\alpha_2\alpha_3\alpha_4\dots}$ 
we obtain that the first  and the last term 
vanishes as can be seen from following
examples 
\begin{eqnarray}
\epsilon^{\alpha_2\alpha_3\alpha_4}
\partial_{\alpha_2}\partial_{\alpha_3}A_{\alpha_4}=
-\epsilon^{\alpha_3\alpha_2\alpha_4}
\partial_{\alpha_3}\partial_{\alpha_2}A_{\alpha_4} \ ,
\nonumber \\
\epsilon^{\alpha_2\alpha_3\alpha_4}
\partial_{\alpha_2}\partial_{\alpha_3}X^I=
-\epsilon^{\alpha_3\alpha_2\alpha_4}
\partial_{\alpha_3}\partial_{\alpha_2}X^I \ . 
\nonumber \\
\end{eqnarray}

If we now combine 
 (\ref{pt1f}) with (\ref{pt2})
we  obtain that the 
tachyon current  has
natural interpretation as the current
for the scalar mode $t(\xi)$ 
that parametrises location of
D(p-1)-brane in the $x^9$ direction
\begin{equation}\label{Jtf}
J_T=-V\tilde{J}_9 \  
\end{equation}
with $\tilde{J}_9$ given in
(\ref{currentAbps}).

Finally we will  analyse 
currents  
$J_K$ given in (\ref{currentXK}).
Let us start with the first term
in (\ref{currentXK}) 
\begin{eqnarray}\label{jkp1fp}
\sum_{n\leq 0}
\frac{1}{n!(2!)^nq!}
\epsilon^{\mu_1\dots \mu_{p+1}}
V(T)\partial_Kb_{MN}\partial_{\mu_1}X^M
\partial_{\mu_2}X^N(\mF)^{n-1}_{\mu_3
\dots \mu_{2n}}
C_{\mu_{2n+1}\dots \mu_p}\partial_{\mu_{p+1}}
T=\nonumber \\
aVf'\sum_{n\leq 0}
\frac{1}{n!(2!)^nq!}\epsilon^{\alpha_1
\dots\alpha_p x}\left(
\partial_Kb_{MN}\partial_{\alpha_1}Y^I
\partial_{\alpha_2}
Y^N 
(\mF)^{n-1}_{\alpha_3\dots \alpha_{2n}}
C_{\alpha_{2n+1}\dots\alpha_p}+
\right.
\nonumber \\
\left.+2(n-1)\partial_Kb_{MN}\partial_{\alpha_1}Y^M
\partial_{\alpha_2}
Y^N b_{9\alpha_4}\partial_{\alpha_3}t
(\mF)^{n-2}_{\alpha_5\dots\alpha_{2n}}
\tilde{C}_{\alpha_{2n+1}\dots\alpha_p} 
\right. \nonumber 
\\
\left.
+q\partial_Kb_{MN}\partial_{\alpha_1}Y^M
\partial_{\alpha_2}
Y^N (\tilde{\mF})^{n-1}_{\alpha_3\dots\alpha_{2n}}
C_{x\alpha_{2n+2}\dots\alpha_p}
\partial_{\alpha_{2n+1}}t
 \right) \ . 
\nonumber \\
\end{eqnarray}
In the previous expressions 
we have included
the terms with tilde from the same reasons as
was argued in case of gauge 
field current.  We can also simplify the
expression above using the fact that
\begin{equation}
\epsilon^{\alpha_1\alpha_2
\alpha_3\dots\alpha_{2n}\dots}
\left(\mF^{n-1}_{\alpha_3\dots\alpha_{2n}}
+2(n-1)b_{9I}\partial_{\alpha_3}
t\partial_{\alpha_4}X^I(\mF)^{n-2}_{\alpha_5
\dots\alpha_{2n}}\right)=
\epsilon^{\alpha_1\alpha_2
\alpha_3\dots\alpha_{2n}\dots}
(\tilde{\mF})^{n-1}_{\alpha_3\dots\alpha_{2n}} \ .
\end{equation} 
In the same way we can see that the
last term in (\ref{jkp1fp}) combine with
the first term in (\ref{jkp1fp}) so that
we can replace $C_{\alpha_{2n+1}\dots \alpha_p}$
with $\tilde{C}_{\alpha_{2n+1}\dots \alpha_p}$.
Then (\ref{jkp1fp}) can be written
as 
\begin{eqnarray}\label{jkp1f}
\nonumber \\
aVf'\sum_{n\leq 0}
\frac{1}{n!(2!)^nq!}\epsilon^{\alpha_1
\dots\alpha_p x}\left(
\partial_Kb_{MN}\partial_{\alpha_1}Y^M
\partial_{\alpha_2}Y^N 
(\tilde{\mF})^{n-1}_{\alpha_3\dots \alpha_{2n}}
\tilde{C}_{\alpha_{2n+1}\dots\alpha_p}
\right)
\ .
\nonumber \\
\end{eqnarray}
Looking on the form of the expression
on the second line in (\ref{currentXK}) it is clear
that it can be analysed in the same way as we
did above 
\begin{eqnarray}\label{jkp2fp}
\epsilon^{\mu_1\dots\mu_{p+1}}
V(T)(\mF)^n_{\mu_1\dots \mu_{2n}}
\partial_K C_{M_1\dots M_q}
\partial_{\mu_{2n+1}}X^{M_1}\dots 
\partial_{\mu_p}X^{M_q}\partial_{\mu_{p+1}}T=
 \nonumber \\
=Vaf'\epsilon^{\alpha_1\dots\alpha_px}
\left(
(\mF)^n_{\alpha_1\dots\alpha_{2n}}
\partial_KC_{\alpha_{2n+1}\dots\alpha_p}+\right. 
\nonumber \\
\left.+2nb_{9M}\partial_{\alpha_1}t
\partial_{\alpha_2}Y^M
(\tilde{\mF})^{n-1}_{\alpha_3\dots\alpha_{2n}}
\partial_K\tilde{C}_{\alpha_{2n+1}\dots\alpha_p}+
\right. \nonumber \\
\left. q\epsilon^{\alpha_1\dots \alpha_px}
(\tilde{\mF})^n_{\alpha_1\dots\alpha_{2n}}
\partial_KC_{x\alpha_{2n+2}\dots \alpha_p}
\partial_{\alpha_{2n+1}}t \right) \  \nonumber \\
\end{eqnarray}
that using the same arguments as were
given below (\ref{jkp1fp}) it can be
rewritten in more suggestive form
\begin{equation}\label{jkp2f}
aVf'\epsilon^{\alpha_1\dots\alpha_p x}
(\tilde{\mF})^n_{\alpha_1\dots\alpha_{2n}}
\partial_K \tilde{C}_{\alpha_{2n+1}\dots\alpha_p}
 \ . 
\end{equation}
Now let us consider expression
on the third line in (\ref{currentXK})
\begin{eqnarray}\label{jkp3fp}
-2n\epsilon^{\mu_1
\dots\mu_{p+1}}\partial_{\mu_1}
\left[V(T)b_{KM}\partial_{\mu_2}
X^M(\mF)^{n-1}_{\mu_3\dots 
\mu_{2n}}
C_{\mu_{2n+1}\dots \mu_p}\partial_{\mu_{p+1}}T
\right]=
\nonumber \\
=-2naf'V\epsilon^{\alpha_1\dots\alpha_p x}
\partial_{\alpha_1}
\left[b_{KM}\partial_{\alpha_2}Y^M
(\mF)^{n-1}_{\alpha_3
\dots\alpha_{2n}}
C_{\alpha_{2n+1}\dots 
\alpha_p}\right] - \nonumber \\
-2nVaf'\partial_x[b_{KM}
(\tilde{\mF})^{n-1}_{\alpha_2
\dots\alpha_{2n}}
\tilde{C}_{\alpha_{2n+1}
\dots \alpha_p}]\partial_{\alpha_1} t
\partial_{\alpha_2}Y^M 
\nonumber \\
-4n(n-1)
af'V\epsilon^{\alpha_1\dots\alpha_px}
\partial_{\alpha_1}\left[
b_{KM}
\partial_{\alpha_2}
Y^Jb_{9N}\partial_{\alpha_3}t
\partial_{\alpha_4}Y^N
(\tilde{\mF})^{n-2}_{\alpha_5\dots\alpha_{2n}}
\tilde{C}_{\alpha_{2n+1}\dots \alpha_p}
\right]-
\nonumber \\
-2nqaf'V\epsilon^{\alpha_1\dots\alpha_px} 
\partial_{\alpha_1}
\left[b_{KM}\partial_{\alpha_2}Y^M
\partial_{\alpha_3} t
(\tilde{\mF})^{n-1}_{\alpha_3\dots\alpha_{2n}}
C_{x\alpha_{2n+2}\dots \alpha_p}\partial_{\alpha_{2n+1}}t
\right]  \ , \nonumber\\ 
\end{eqnarray}
where we have used the fact that
\begin{equation}
\epsilon^{\alpha_1\alpha_2\dots}
\partial_{\alpha_1}[Vaf']\partial_{\alpha_2}t=
-\epsilon^{\alpha_1\alpha_2\dots}
\partial_x[Vaf']\partial_{\alpha_1}t\partial_{\alpha_2}t=
0 \ . 
\end{equation}
We see that the expression on the 
first line in (\ref{jkp3fp}) together with
the expression on the third and the fourth line
arise from 
 expansions of $\tilde{\mF}$ and $\tilde{C}$ 
in terms of  $\mF\ , C$ and $\partial t$.
Consequently 
(\ref{jkp3fp}) takes the form
\begin{eqnarray}\label{jkp3f}
-2naf'V\epsilon^{\alpha_1\dots\alpha_p x}
\partial_{\alpha_1}
\left[b_{KM}\partial_{\alpha_2}Y^M
(\tilde{\mF})^{n-1}_{\alpha_3
\dots\alpha_{2n}}
\tilde{C}_{\alpha_{2n+1}\dots 
\alpha_p}\right] - \nonumber \\
-2nVaf'\partial_x[b_{KM}
(\tilde{\mF})^{n-1}_{\alpha_2
\dots\alpha_{2n}}
\tilde{C}_{\alpha_{2n+1}
\dots \alpha_p}]\partial_{\alpha_1} t
\partial_{\alpha_2}Y^M \ .  
\nonumber \\
\end{eqnarray}
Finally, we will analyse the expression
on the fourth line in (\ref{currentXK})
\begin{eqnarray}\label{jkp4fp}
-q\epsilon^{\mu_1
\dots\mu_{p+1}}\partial_{\mu_{2n+1}}
\left[V(T)(\mF)^{n}_{\mu_1\dots \mu_{2n}}
C_{KM_2\dots M_q}\partial_{\mu_{2n+2}}
X^{M_2}\dots \partial_{\mu_p}X^{M_q}
\partial_{\mu_{p+1}}T\right]= \nonumber \\
=-qaVf'\epsilon^{\alpha_1\dots \alpha_px}
\partial_{\alpha_{2n+1}}
\left[(\mF)^n_{\alpha_1\dots\alpha_{2n}}
C_{K\alpha_{2n+2}\dots\alpha_p}\right]+\nonumber \\
-qaVf'\epsilon^{\alpha_1\alpha_px}
\partial_x\left[(\tilde{\mF})^n_{\alpha_1\dots\alpha_{2n}}
\tilde{C}_{K\alpha_{2n+2}
\dots\alpha_p}\right]\partial_{\alpha_{2n+1}}
t-\nonumber \\
-2nqaVf'\epsilon^{\alpha_1\dots\alpha_px}
\partial_{\alpha_{2n+1}}
\left[b_{9M}\partial_{\alpha_1}t\partial_{\alpha_2}Y^M
(\tilde{\mF})^{n-1}_{\alpha_3\dots\alpha_{2n}}
\tilde{C}_{K\alpha_{2n+2}\dots\alpha_p}\right]-\nonumber \\
-q(q-1)\epsilon^{\alpha_1\alpha_p x}
\partial_{\alpha_{2n+1}}
\left[(\tilde{\mF})^n_{\alpha_1\dots\alpha_{2n}}
C_{Kx\alpha_{2n+3}\dots\alpha_p}\partial_{\alpha_2}
t\right]  \ . \nonumber \\
\end{eqnarray}
Following discussion given below
(\ref{jkp3fp}) we can rewrite
(\ref{jkp4fp}) into the form 
\begin{eqnarray}\label{jkp4f}
-qaVf'\epsilon^{\alpha_1\dots \alpha_px}
\partial_{\alpha_{2n+1}}
\left[(\tilde{\mF})^n_{\alpha_1\dots\alpha_{2n}}
\tilde{C}_{K\alpha_{2n+2}\dots\alpha_p}\right]
+\nonumber \\
-qaVf'\epsilon^{\alpha_1\alpha_px}
\partial_x\left[(\tilde{\mF})^n_{\alpha_1\dots\alpha_{2n}}
\tilde{C}_{K\alpha_{2n+2}\dots\alpha_p}\right]
\partial_{\alpha_{2n+1}}
t \ . \nonumber \\
\end{eqnarray}
If we look on the expressions
 in (\ref{jkp3f}) and (\ref{jkp4f}) 
 we  see that there are two terms
\begin{eqnarray}
\nonumber \\
-2nVaf'\partial_x[b_{KM}
(\tilde{\mF})^{n-1}_{\alpha_2
\dots\alpha_{2n}}
\tilde{C}_{\alpha_{2n+2}
\dots \alpha_p}]\partial_{\alpha_1} t
\partial_{\alpha_2}Y^M \  ,\nonumber \\
-qaVf'\epsilon^{\alpha_1\alpha_px}
\partial_x\left[(\tilde{\mF})^n_{\alpha_1\dots\alpha_{2n}}
\tilde{C}_{K\alpha_{2n+2}\dots\alpha_p}
\right]\partial_{\alpha_{2n+1}}t \ . 
\end{eqnarray}
These terms are again important for
an interpretation of $t(\xi)$ as an embedding
coordinate as was more carefully discussed
above. 
Finally collecting (\ref{jkp1f}) , (\ref{jkp2f}),
(\ref{jkp3f}) and (\ref{jkp4f}) 
together we obtain
that the current $J_K$ 
takes the form
 \begin{equation}\label{jkf}
J_K=af'V\tilde{J}_K \ , 
\end{equation}
where $\tilde{J}_K$ is given in
(\ref{currentxkbps}).

As the final point we should determine
the form of the current $J_9$. 
In fact, since in the analysis performed
above there is nothing special about the
index $K$ it is clear that the result obtained
there can be applied for $K=9$ as well
and we get
\begin{equation}\label{jkxf}
J_x=af'V\tilde{J}_x \ .
\end{equation}

{\bf Acknowledgement}

This work was supported by the
Czech Ministry of Education under Contract No.
MSM 0021622409.



\begin{thebibliography}{20}



\bibitem{Sen:1999mg}
  A.~Sen,
\emph{ ``Non-BPS states 
and branes in string theory,''}
  
  arXiv:hep-th/9904207.

\bibitem{Ohmori:2001am}
  K.~Ohmori,
\emph{ ``A review on tachyon 
condensation in 
open string field theories,''}

  arXiv:hep-th/0102085.

\bibitem{Taylor:2002uv}
  W.~Taylor,
   \emph{``Lectures on D-branes, 
tachyon condensation, 
and string field theory,''}

  arXiv:hep-th/0301094.

\bibitem{Taylor:2003gn}
  W.~Taylor and B.~Zwiebach,
  \emph{ ``D-branes, tachyons, 
and string field theory,''}
  arXiv:hep-th/0311017.

\bibitem{Sen:2004nf}
  A.~Sen,
 \emph{``Tachyon dynamics 
in open string theory,''}
  arXiv:hep-th/0410103.

\bibitem{Sen:1999md}  
A.~Sen,
  \emph{``Supersymmetric world-volume action 
for non-BPS D-branes,''}
  JHEP {\bf 9910} (1999) 008
  [arXiv:hep-th/9909062].

\bibitem{Bergshoeff:2000dq}
  E.~A.~Bergshoeff, M.~de Roo, 
T.~C.~de Wit, E.~Eyras and S.~Panda,
\emph{``T-duality and actions for non-BPS D-branes,''}
  JHEP {\bf 0005} (2000) 009
  [arXiv:hep-th/0003221].

\bibitem{Garousi:2000tr}
  M.~R.~Garousi,
 \emph{ ``Tachyon couplings on 
non-BPS D-branes and Dirac-Born-Infeld action,''}
  Nucl.\ Phys.\ B {\bf 584} (2000) 284
  [arXiv:hep-th/0003122].

\bibitem{Kluson:2000iy}
  J.~Kluson,
\emph{ ``Proposal for non-BPS D-brane action,''}
  Phys.\ Rev.\ D {\bf 62} (2000) 126003
  [arXiv:hep-th/0004106].  
HEP-TH 0004106;




\bibitem{Sen:2002qa}
  A.~Sen,
  \emph{``Time and tachyon,''}
  Int.\ J.\ Mod.\ Phys.\ A {\bf 18} 
(2003) 4869
  [arXiv:hep-th/0209122].

\bibitem{Fotopoulos:2003yt}
  A.~Fotopoulos and A.~A.~Tseytlin,
  \emph{``On open superstring 
partition function in 
inhomogeneous rolling tachyon
  background,''}
  JHEP {\bf 0312}, 025 (2003)
  [arXiv:hep-th/0310253].

\bibitem{Sen:2003bc}
  A.~Sen,
  \emph{``Open and closed 
strings from unstable D-branes,''}
  Phys.\ Rev.\ D {\bf 68}, 106003 (2003)
  [arXiv:hep-th/0305011].

\bibitem{Kutasov:2003er}
  D.~Kutasov and V.~Niarchos,
  \emph{``Tachyon effective actions 
in open string theory,''}
  Nucl.\ Phys.\ B {\bf 666}, 56 (2003)
  [arXiv:hep-th/0304045].

\bibitem{Niarchos:2004rw}
  V.~Niarchos,
  \emph{``Notes on tachyon 
effective actions and 
Veneziano amplitudes,''}
  Phys.\ Rev.\ D {\bf 69}, 106009 (2004)
  [arXiv:hep-th/0401066].



\bibitem{Sen:2003tm}
  A.~Sen,
  \emph{``Dirac-Born-Infeld 
action on the tachyon kink and vortex,''}
  Phys.\ Rev.\ D {\bf 68} (2003) 066008
  [arXiv:hep-th/0303057].

\bibitem{Kim:2005he}
  C.~Kim, Y.~Kim, H.~h.~Kwon and O.~K.~Kwon,
  \emph{``BPS D-branes from an 
unstable D-brane in a curved background,''}
  arXiv:hep-th/0504130.

\bibitem{Kim:2003in}
  C.~j.~Kim, Y.~b.~Kim and C.~O.~Lee,
\emph{``Tachyon kinks,''}
  JHEP {\bf 0305} (2003) 020
  [arXiv:hep-th/0304180].

\bibitem{Kim:2003ma}
  C.~Kim, Y.~Kim, O.~K.~Kwon and C.~O.~Lee,
\emph{``Tachyon kinks on unstable Dp-branes,''}
  JHEP {\bf 0311} (2003) 034
  [arXiv:hep-th/0305092].

\bibitem{Banerjee:2004cw}
  R.~Banerjee, Y.~Kim and O.~K.~Kwon,
 \emph{``Noncommutative tachyon 
kinks as D(p-1)-branes from unstable 
D p-brane,''}
  JHEP {\bf 0501} (2005) 023
  [arXiv:hep-th/0407229].

\bibitem{Bazeia:2004vc}
  D.~Bazeia, R.~Menezes and J.~G.~Ramos,
 \emph{``Regular and 
periodic tachyon kinks,''}
  Mod.\ Phys.\ Lett.\ A {\bf 20}
 (2005) 467
  [arXiv:hep-th/0401195].

\bibitem{Copeland:2003df}
  E.~J.~Copeland, P.~M.~Saffin 
and D.~A.~Steer,
\emph{``Singular tachyon 
kinks from regular profiles,''}
  Phys.\ Rev.\ D {\bf 68} (2003) 065013
  [arXiv:hep-th/0306294].

\bibitem{Brax:2003rs}
  P.~Brax, J.~Mourad and D.~A.~Steer,
  \emph{``Tachyon kinks on non BPS D-branes,''}
  Phys.\ Lett.\ B {\bf 575} (2003) 115
  [arXiv:hep-th/0304197].


\bibitem{Kluson:2004yk}
  J.~Kluson,
\emph{``Non-BPS Dp-brane 
in the background of 
NS5-branes on transverse R**3 x
S**1,''}
  JHEP {\bf 0503} (2005) 032
  [arXiv:hep-th/0411014].

\bibitem{Kluson:2005qx}
  J.~Kluson,
  \emph{``Non-BPS Dp-brane 
in Dk-brane background,''}
  JHEP {\bf 0503} (2005) 044
  [arXiv:hep-th/0501010].

\bibitem{Okuyama:2003wm}
  K.~Okuyama,
  \emph{``Wess-Zumino term 
in tachyon effective action,''}
  JHEP {\bf 0305} (2003) 005
  [arXiv:hep-th/0304108].


\bibitem{Kraus:2000nj}
  P.~Kraus and F.~Larsen,
  \emph{``Boundary string 
field theory of the DD-bar system,''}
  Phys.\ Rev.\ D {\bf 63} (2001) 106004
  [arXiv:hep-th/0012198].

\bibitem{Kennedy:1999nn}
  C.~Kennedy and A.~Wilkins,
\emph{``Ramond-Ramond couplings on 
brane-antibrane systems,''}
  Phys.\ Lett.\ B {\bf 464} (1999) 206
  [arXiv:hep-th/9905195].

\bibitem{Billo:1999tv}
  M.~Billo, B.~Craps and F.~Roose,
\emph{ ``Ramond-Ramond couplings of non-BPS D-branes,''}
  JHEP {\bf 9906} (1999) 033
  [arXiv:hep-th/9905157].

\bibitem{Takayanagi:2000rz}
  T.~Takayanagi, S.~Terashima and T.~Uesugi,
  \emph{``Brane-antibrane action 
from boundary string field theory,''}
  JHEP {\bf 0103} (2001) 019
  [arXiv:hep-th/0012210].

\bibitem{Kluson:2005hd}
  J.~Kluson,
  \emph{``Note about tachyon 
kink in nontrivial background,''}
  arXiv:hep-th/0506250.

\bibitem{Skenderis:2002vf}
  K.~Skenderis and M.~Taylor,
 \emph{``Branes in AdS and pp-wave spacetimes,''}
  JHEP {\bf 0206} (2002) 025
  [arXiv:hep-th/0204054].





\end{thebibliography}
\end{document}